\documentclass[useAMS,usenatbib]{mn2e}
\usepackage {graphicx}
\usepackage{multirow}
\usepackage{subfig}
\usepackage{float}
\usepackage{amsmath}
\usepackage{subfloat}
\usepackage{afterpage}
\usepackage{rotating}
\usepackage{caption}

\usepackage{aasmacros}
\usepackage{booktabs}
\usepackage{courier}

\newcommand {\aple} {\ {\raise-.5ex\hbox{$\buildrel<\over\sim$}}\ }
\newcommand {\solmass}{\textrm{M}_{\odot}}
\newcommand {\Mvir}{\textrm{M}_{\textrm{vir}}}

\title[ISM studies towards HESS\,J1825$-$137]{ ISM gas studies towards the TeV PWN HESS J1825-137 and northern region.}
\author[F.Voisin, G.Rowell, M.G Burton, A.Walsh, Y.Fukui, F. Aharonian]{F.Voisin$^{1}$\thanks{E-mail:
fabien.voisin@adelaide.edu.au (AVR); fabien.voisin@adelaide.edu.au(ANO)}, 
G.Rowell$^{1}$, M.G.Burton$^{2}$, A.Walsh$^{3}$, Y.Fukui$^{4}$, F.Aharonian$^{5,6}$\\
$^{1}$ School of Physical Science, Adelaide university, North Terrace , Adelaide , SA 5005, Australia\\
$^{2}$ School of Physics, University of New South Wales, Sydney, NSW 2052, Australia\\
$^{3}$ International Centre for Radio Astronomy Research, Curtin University, GPO Box U1987, Perth WA 6845, Australia\\
$^{4}$ Department of Physics, University of Nagoya, Furo-cho, Chikusa-ku, Nagoya, 464-8601, Japan\\
$^{5}$ Max-Planck-Institut f\"ur Kernphysik, P.O. Box 103980, D-69029 Heidelberg, Germany\\
$^{6}$ Dublin Institute for Advanced Studies, 31 Fitzwilliam Place, Dublin 2, Ireland\\}

\begin{document}
\label{firstpage}
\pagerange{\pageref{firstpage}--\pageref{lastpage}} \pubyear{2016}

\maketitle

\begin{abstract}
HESS\,J1825$-$137 is a  pulsar wind nebula (PWN) whose TeV emission extends across $\sim$ 1 $\deg$.  
Its large asymmetric shape indicates that its progenitor supernova interacted with a molecular cloud located in the north of the PWN as detected by previous 
CO Galactic survey (e.g \citealt{Lem2006}).

Here we provide a detailed picture of the ISM towards the region north of HESS\,J1825$-$137, 
 with the analysis of the dense molecular gas from our 7mm and 12mm Mopra survey and the more diffuse molecular gas from the Nanten CO(1--0) and GRS $^{13}$CO(1--0) surveys. 
Our focus is the possible association between HESS\,J1825$-$137 and the unidentified TeV source to the north, HESS\,J1826$-$130.
We report several dense molecular regions whose kinematic distance matched the dispersion measured distance of the pulsar. 
Among them, the dense molecular gas located at \mbox{(RA, Dec)=(18.421h,$-13.282^{\circ}$)} shows enhanced turbulence and we suggest that the velocity structure in this region may  be explained by a cloud-cloud collision scenario.

Furthermore, the presence of a H$\alpha$ rim may be the first evidence of the progenitor SNR of the pulsar PSR\,J1826--1334 
as the distance between the H$\alpha$ rim and the TeV source matched with the predicted SNR radius \mbox{$R_{\text{SNR}}\sim120$\,pc}. 

From our ISM study, we identify a few plausible origins of the HESS\,J1826$-$130 emission, including the progenitor SNR of PSR\,J1826$-$1334 and the PWN\,G018.5$-$0.4 powered by PSR\,J1826$-$1256.
A deeper TeV study however, is required to fully identify the origin of this mysterious TeV source.

\end{abstract}


\section{Introduction}

HESS\,J1825$-$137 is one of the brightest and most extensive pulsar wind %
nebulae (PWNe) detected in TeV $\gamma$-rays \citep{Aha2006}. %
It is powered by the high spin-down power \mbox{($\dot{E}_{\text{SD}}=2.8\times 10^{36}$\,erg/s)} 
pulsar PSR\,J1826$-$1334 with a dispersion measure distance of \mbox{3.9$\pm$0.4\,kpc}
and characteristic age $\tau_c \sim 20$\,kyr.

PWNe represent a significant fraction of the Galactic TeV $\gamma$-ray source %
population. They convert a varying fraction
of their pulsars' spin down energy $\dot{E}_\text{SD}$ into high energy electrons. %
The electron flow is temporally randomized and re-accelerated at a 
termination shock resulting from pressure from the surrounding interstellar
medium (ISM). Inverse-Compton (IC)
up-scattering of soft photons then leads to TeV $\gamma$-rays, %
and associated synchrotron radio to X-ray emission. %

The morphology of PWNe can be heavily influenced by the ISM. The 
interaction of the progenitor supernova shock with adjacent molecular 
clouds can  lead to a reverse shock propagating back into 
the PWN \citep{Blon2001}, giving rise to an asymmetry in the radio, 
X-ray and $\gamma$-ray emission that can trail away from the pulsar along 
the pulsar-molecular cloud axis.

HESS\,J1825$-$137 is an excellent example of this situation. The morphology
of HESS\,J1825$-$137 (Fig. 1) displays a clear asymmetry 
with respect to PSR\,J1826$-$1334, and a molecular cloud to the north revealed 
by \citet{Lem2006}, with the bulk of the TeV $\gamma$-ray 
extending up to a degree south of the pulsar.

\begin{figure*}
\hspace{-1.5cm}
\includegraphics [trim=0 0 0 0,clip,width=0.93\textwidth,angle=0]{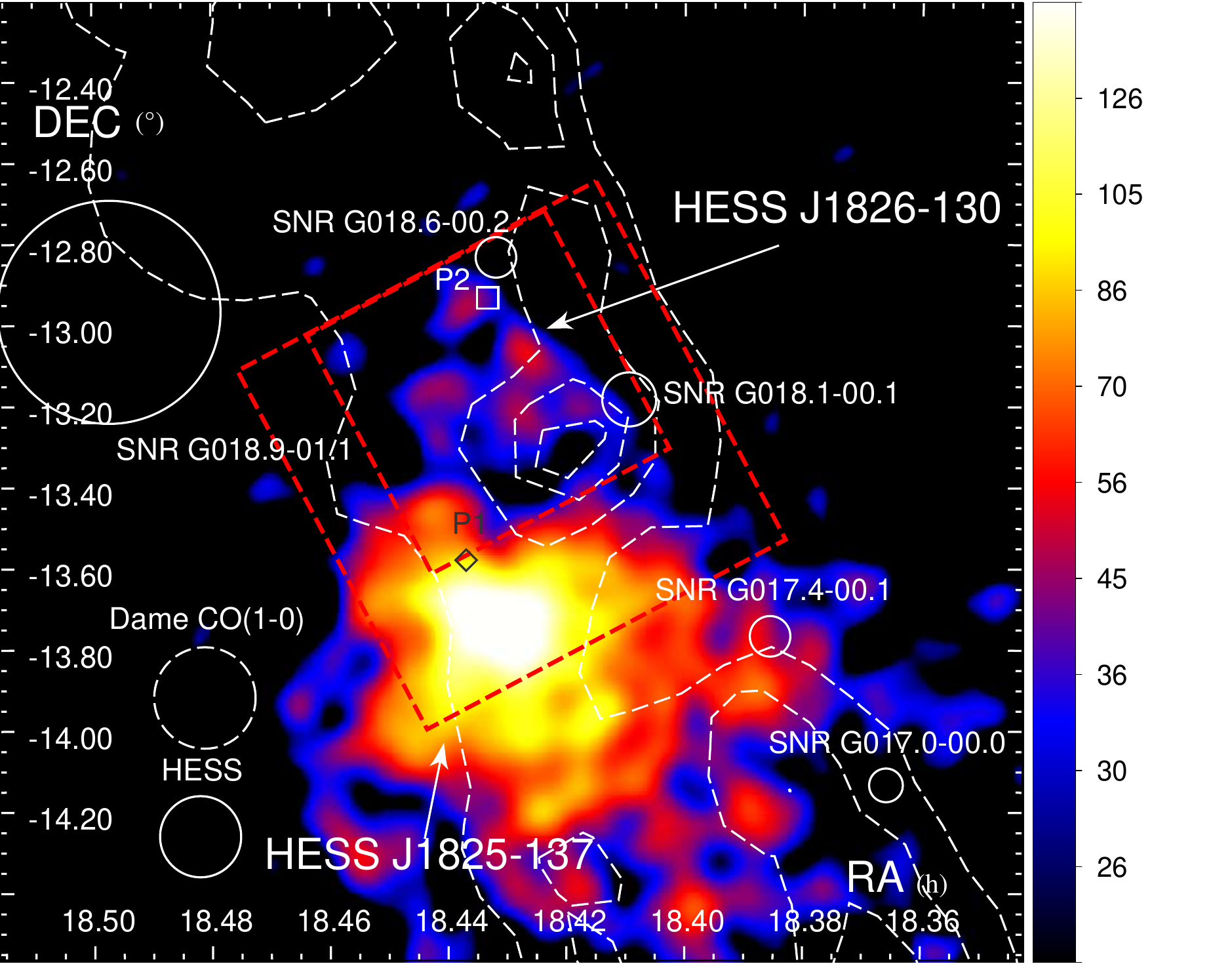}
\hspace{-5cm}
\vspace{1cm}
\caption{ HESS excess counts image ($>100$ GeV) towards HESS\,J1825$-$137 overlaid by the CO(1--0) integrated intensity contour (40, 60, 80, 100 K km/s) between $v_{\text{lsr}}=40-60$\,km/s from \citet{Dame2001} as revealed by \citet{Lem2006}.
The white circles represent the different SNRs detected \citep{Brogan2006}. P1 indicates the pulsar PSR\,J1826-1334's location while P2 shows the position of PSR\,J1826--1256. The small and large red dashed boxes (see online version)
represent our 7mm and 12mm Mopra coverage respectively.}
\label{NantenCS}

\end{figure*}

Interestingly, the weak TeV $\gamma$-ray emission component to the north labeled HESS\,J1826$-$130 \citep{Deil2015} appears to spatially overlap
this northern molecular cloud (see Fig. \ref{NantenCS}). Such an overlap could result from the interaction
of multi-TeV cosmic-rays with molecular clouds, and thus raises the possibility of
cosmic-ray acceleration in the vicinity, notably from HESS\,J1825-137's progenitor SNR. 

We also note the presence of two additional SNRs in the region : G018.1$-$0.1 and G018.6$-$0.2 \citep{Brogan2006} as shown in Fig. \ref{NantenCS}.
\citet{Ray1} also discovered the radio quiet pulsar PSR\,J1826-1256 (P2 in Fig. \ref{NantenCS}) with a spin down luminosity \mbox{$\dot{E}_{\text{SD}}=3.6\times10^{36}$ erg s$^{-1}$}, a period $P=100$\,ms 
and a characteristic age $\tau_c=13$\,kyr and powering the PWN\,G018.5--0.4 observed in X-rays by \citet{Robert} . Although no dispersion measure could be derived, \citet{Wang} argues that the pulsar is located at a distance $d=1.2-1.4$\,kpc.
Consequently, it is possible each of these sources could also contribute to the TeV emission inside HESS\,J1826$-$130.

To further investigate this issue, higher resolution mapping (improving that 
of the 8$^\prime$ resolution of the CO(1--0) data used by \citealt{Lem2006})
of the molecular cloud and surrounding region is needed to probe the density and dynamics of
the gas. In this paper, we made use of CO, and $^{13}$CO survey data (from the
Nanten telescope and the Galactic Ring Survey - GRS) plus new mapping
of dense gas tracers such as carbon monosulphide (CS) and ammonia (NH$_3$)
with the Mopra telescope in Australia.

In section \S\ref{section2}, we review the properties of the Mopra and Nanten telescopes, and the GRS as well as the methodology used to reduce our Mopra observations. 
Then, in section \S\ref{section3}, we briefly introduce the different gas tracers that we detected towards HESS\,J1826$-$130 and their physical properties.
We present  the results of our observations and provide gas parameter estimates for various regions using our CO, CS and NH$_3$ analysis in section \S\ref{section4}.
 Finally, in section \S\ref{section5}, we  discuss the dynamics of the dense gas and finally discuss a few possible counterparts to the HESS\,J1826$-$130 emission.

\section[]{Data observation and reduction}
\label{section2}
\subsection{Mopra}
We observed a 60$^{\prime}\times60^{\prime}$ region (large red dashed box in Fig. \ref{NantenCS}) centred at \mbox{(RA,Dec)=(18.425h,$-13.3^{\circ}$)} in the 12mm band from the 19$^{\text{th}}$ to the 27$^{\text{th}}$ of January 2012 with the 22 metre Mopra
telescope. We combined our observations with the HOPS survey \citep{Walsh}, which covered the Galactic plane within the Galactic latitude b=-0.5 to b=0.5 to achieve better sensitivity (see Table 1).
Additionally, we observed a 40$^{\prime}\times40^{\prime}$ subsection (small red dashed box in Fig. \ref{NantenCS}) centred at \mbox{(RA,Dec)=(18.427h,$-13.17^{\circ}$)} in the 7mm band.
Four 7mm ON-OFF switched deep pointings were taken at \mbox{(RA,Dec)=(18.419h,$-13.38^{\circ}$)}, \mbox{(18.423h,$-13.34^{\circ}$)}, \mbox{(18.419h,$-13.31^{\circ}$)} and \mbox{(18.420h,$-13.27^{\circ}$)} from the 13$^{\text{th}}$ until the 21$^{\text{st}}$ of April to search for
 additional emission from the isotopologue C$^{34}$S(1--0). Unfortunately, our observations were not sensitive enough to detect such emission.
Finally, another 7mm deep pointing was taken in August 2014 towards the position \mbox{(RA,Dec)=(18.419h,$-14.04^{\circ}$)} towards a molecular cloud south of HESS\,J1825$-$137.

For these observations, we used the Mopra spectrometer MOPS in `zoom' mode, which allowed the recording of sixteen sub-bands, each consisting of 4096 channels and a 137.5 MHz bandwidth, simultaneously.
The Mopra OTF mapping fully sampled the region with a beam size of 2$^{\prime}$ (12mm) and 1$^{\prime}$ (7mm), a velocity resolution of 0.4 km/s (12mm) and 0.2 km/s (7mm).
For the reduction of our OTF observations, we first used \texttt{Livedata}\footnote{http://www.atnf.csiro.au/computing/software/livedata/} which outputs the spectra of each scan using an OFF position as reference. We used a first-order polynomial fit to
subtract the baseline.
 Then, we used \texttt{Gridzilla} \footnote{http://www.atnf.csiro.au/computing/software/Gridzilla/} to produce 3D cubes of each sub-bands in antenna temperature $T_{A}^{*}$ (K) as a function of RA, Dec and line of sight velocity. We used a 15$^{\prime\prime}$ 
 grid to map our region and the data were finally smoothed via a Gaussian with 1.25$^{\prime}$ FWHM in order to smooth out fluctuations. The $T_{\text{rms}}$/channel 
 of each map in which detections have been found are listed in Table \ref{list1}.

Finally, we used the  \texttt{ASAP}\footnote{http://svn.atnf.csiro.au/trac/asap} software to reduce our ON-OFF deep pointing observations. The OFF position measurement was used to obtain the antenna temperature $T_{A}^{*}$ 
of each scans. The achieved $T_\text{{rms}}$/channel ranges from 0.05 to 0.1 K.  
The beam temperature T$_{\text{mb}}$  of maps and ON/OFF pointings were obtained using the main beam conversion factor  $\eta_{\text{mb}}$  determined by \citet{Urquhart2010} (see Appendix \ref{app:couplingfactor}).
\subsection{Nanten and GRS}
To probe the more extended diffuse gas, we made use of more recent CO observations. 
The 4 m Nanten telescope carried out a CO(1--0) survey over the Galactic plane with a 2.6$^{\prime}$ beam size and a sampling grid of 4$^{\prime}$, a velocity resolution $\Delta v$=1.0 km/s \citep{Fukui} and a 
typical $T_{\text{rms}}$/channel value of $\sim0.35$K.
The Galactic Ring Survey $^{13}$CO(1--0) (GRS) mapped the Galactic ring in our Galaxy using the Five College Radio Astronomy Observatory (FCRAO). It has a  beam FWHM of 44$^{\prime\prime}$, a sampling grid of 22$^{\prime\prime}$, 
a velocity resolution $\Delta v\sim$0.2 km/s and an averaged $T_{\text{rms}}$/channel$\sim 0.36$K \citep{JacksonGRS}.
\section[]{Overview of detected lines}

\label{section3}
\begin{table}
 \centering
 \caption{List of all gas tracers detected during our Mopra 7 and 12 mm observations towards HESS\,J1826$-$130, their respective rest frequencies, and the achieved mapping $T_{\text{rms}}$ of the data cubes where 
these emission are found.}
 \begin{tabular}{ccc}
 \toprule
 Tracer & Frequency (GHz) & $T_{\text{rms}}$ (K/channel)\\
 \midrule
 \multicolumn{3}{c}{7mm}\\
  \midrule
  SiO(1--0,v=2) &42.820582 & 0.05 \\
  SiO(1--0,v=0) &43.423864 & 0.05\\
  CH$_{3}$OH(7$_0$--6$_1$)&44.069476 & 0.06 \\
  H51$\alpha$& 45.453720& 0.06 \\
  HC$_{3}$N(5--4,F=4-3)& 46.247580 & 0.06\\
  C$^{34}$S(1--0)& 48.206946 & 0.08 \\
  CS(1--0)& 48.990957 & 0.08\\
  \midrule
  \multicolumn{3}{c}{12mm}\\
  \midrule
  H69$\alpha$ & 19.591110& 0.05\\
  H65$\alpha$ &23.404280& 0.05\\
  NH$_{3}$(1,1)& 23.594470& 0.05\\
  NH$_{3}$(2,2)& 23.722634& 0.05\\
  NH$_{3}$(3,3)& 23.870127& 0.05\\
  H62$\alpha$ &26.939170& 0.05\\
  \bottomrule
 \end{tabular}
\label{list1}
\end{table}
Table \ref{list1} indicates the various spectral lines detected in our analysis of Mopra data.
The following sections review the properties of the major spectral tracers.

\subsection{Carbon monosulfide CS(1-0) and  isotopologues}

A major 7mm line in our study is the J=1--0 emission from the carbon monosulfide (CS) molecule.
CS is commonly found inside dense cores. Its critical density ,$n_{c}$, at $T_{\text{k}}$=10K  is $\sim 2\times 10^4$ cm$^{-3}$ and thus allows study of dense clumps located inside molecular clouds.
The isotopologues of CS namely C$^{34}$S, and C$^{13}$S are generally assumed to  be optically thin given their abundance ratio [CS]/[C$^{34}$S]$\sim$24 and [CS]/[$^{13}$CS]$\sim$75 based on terrestrial measurements (see e.g \citealt{Frierking}). Their detection
can provide estimates of optical depth and thus robust mass estimates of dense cores.

\subsection{Ammonia NH$_{3}$}
\citet{Ho} outlined the properties of the ammonia molecule. The NH$_{3}$(J,K) structure consists of ladders where J is the total momentum 
and K is the momentum from the quadrupole axis.
Only energy states where J=K are metastable and thus can be populated easily.
NH$_{3}$(1,1) spectra consist of one main line surrounded by four satellite lines whose  expected relative strength compared to the main component is $\sim 25$\%.
However, the ratio depends on optical depth, and so this enables an efficient estimation of the dense molecular gas opacity. 
Finally, the relative strength of the NH$_3$(2,2) and NH$_3$(3,3) satellite lines relative to their main components are $\sim$\,5\% and $\sim$\,3\% respectively and it is therefore unlikely to detect these satellite lines.
\newline
\newline
\newline

 \subsection{SiO(1-0)}
 Silicates can be released from dust grains into the gas phase from the crossing of a weak shock inside the molecular clouds \citep{Schilke,Gusdorf}.
 Silicon monoxide is then produced behind the shock and  the non vibrational SiO(1--0,v=0) emission can be detected.
 Its detection becomes optimal for a shock speed \mbox{$v_{\text{s}}=25-50$\,km/s} and a target density \mbox{$n_{\text{H}}=10^{4}$ cm$^{-3}$} \citep{Gusdorf}.
 Our 7mm settings also enabled a search for SiO(1--0,v=1 to 3), and their  non vibrational isotopologues $^{29}$SiO(1--0,v=0) and $^{30}$SiO(1--0,v=0).
\subsection{Other spectral tracers}

The 12mm and 7mm recombination lines H62$\alpha$, H65$\alpha$ and H69$\alpha$ and H51$\alpha$ indicate ionized gas by UV radiation from nearby stars in H\textsc{ii}  complexes.
Thus, these are important  tracers of star formation and photo-dissociation regions where newborn stars radiate inside molecular clouds.

Cyanoacetylene (HC$_{3}$N) can be detected in warm molecular clouds and can be associated with star forming regions. 
This line emission is typically assumed optically thin and the large number of transitions available in the millimetre band allow the computation of physical parameters of molecular clouds \citep{Morris1976}.

Also, the methanol transition CH$_3$OH(7$_0$-6$_1$) is a class\,I maser and generally traces star formation outflows. 
Weak shocks also tend to release CH$_3$OH from the grain mantle and collisionally pump the molecule from increased interaction with H$_2$ (see \citealt{Voronkov} and references therein).

\section{Results and analysis}
\label{section4}
\begin{figure*}
 \begin{minipage}{\textwidth}
 \centering
   \includegraphics[height=0.85\textheight,angle=0]{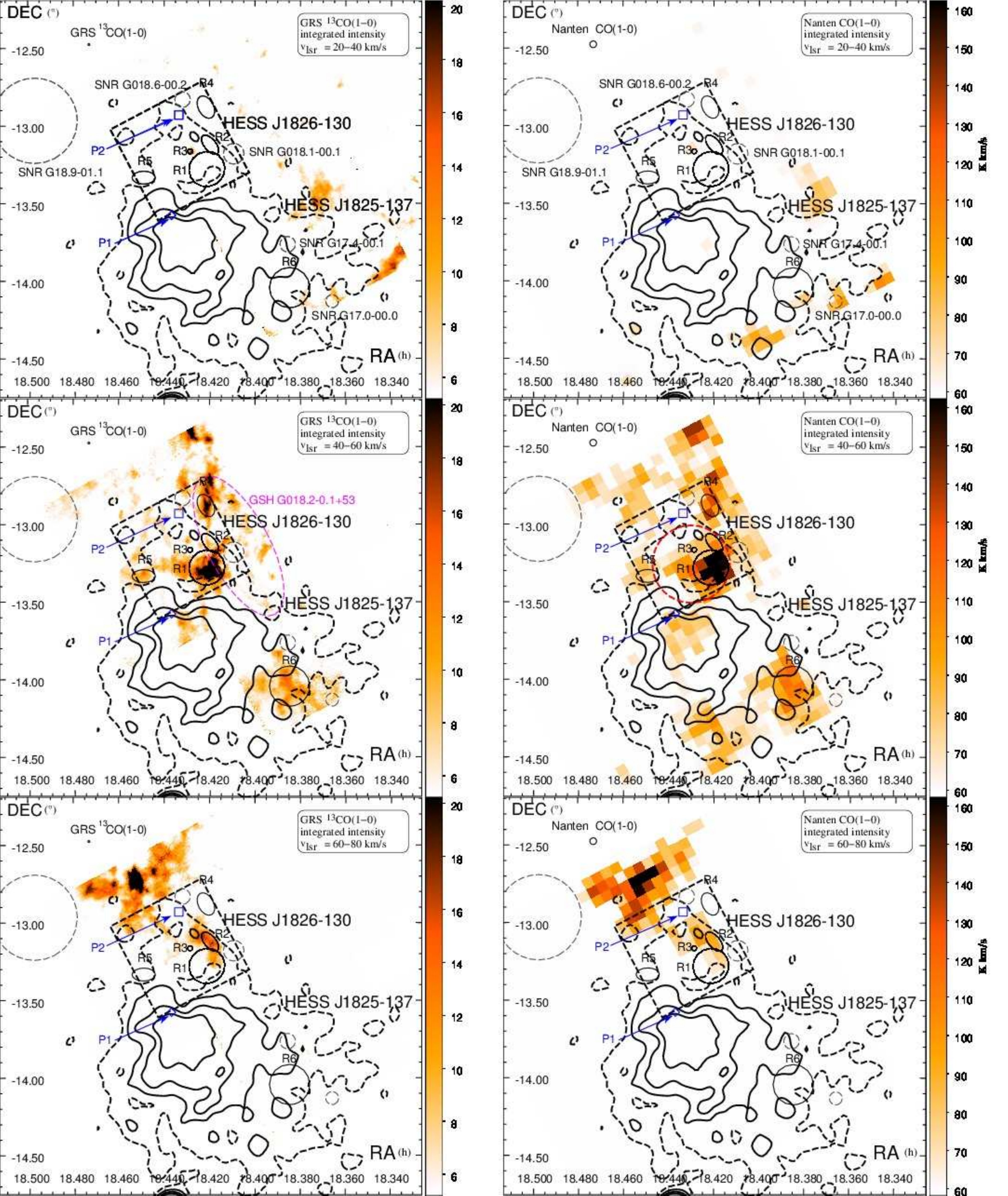}%
  \caption{ GRS $^{13}$CO(1-0) (left) and Nanten CO(1-0) (right) integrated intensity between $v_{\text{lsr}}=20-40$\,km/s, $v_{\text{lsr}}=40-60$\,km/s and $v_{\text{lsr}}=60-80$\,km/s. The different black ellipses represent
  regions for further discussion based on detection of dense gas via the CS(1--0) and NH$_3$(1,1) tracers (see Fig. \ref{CSCOGRS}). The HESS TeV emission from HESS\,J1825$-$137 and HESS\,J1826$-$130 is shown 
  in black contours (dashed and solid) and the surrounding SNRs are displayed in black dashed circles with their label displayed in the first panels. The region covered by our 7mm survey is shown in black dashed box. The 
  putative molecular shell GSH\,18.1-0.2+53 \citep{Paron2013} is shown in purple dashed ellipse (see online version) in the middle-left panel. Finally, the red dashed circle in the middle right panel represents the region whose 
  mass and density have been calculated (see section 4.2). }
 \label{dissociation}
 \end{minipage}
\end{figure*}

\begin{figure*}
 \begin{minipage}{\textwidth}
 \centering
   \includegraphics[height=0.85\textheight,angle=0]{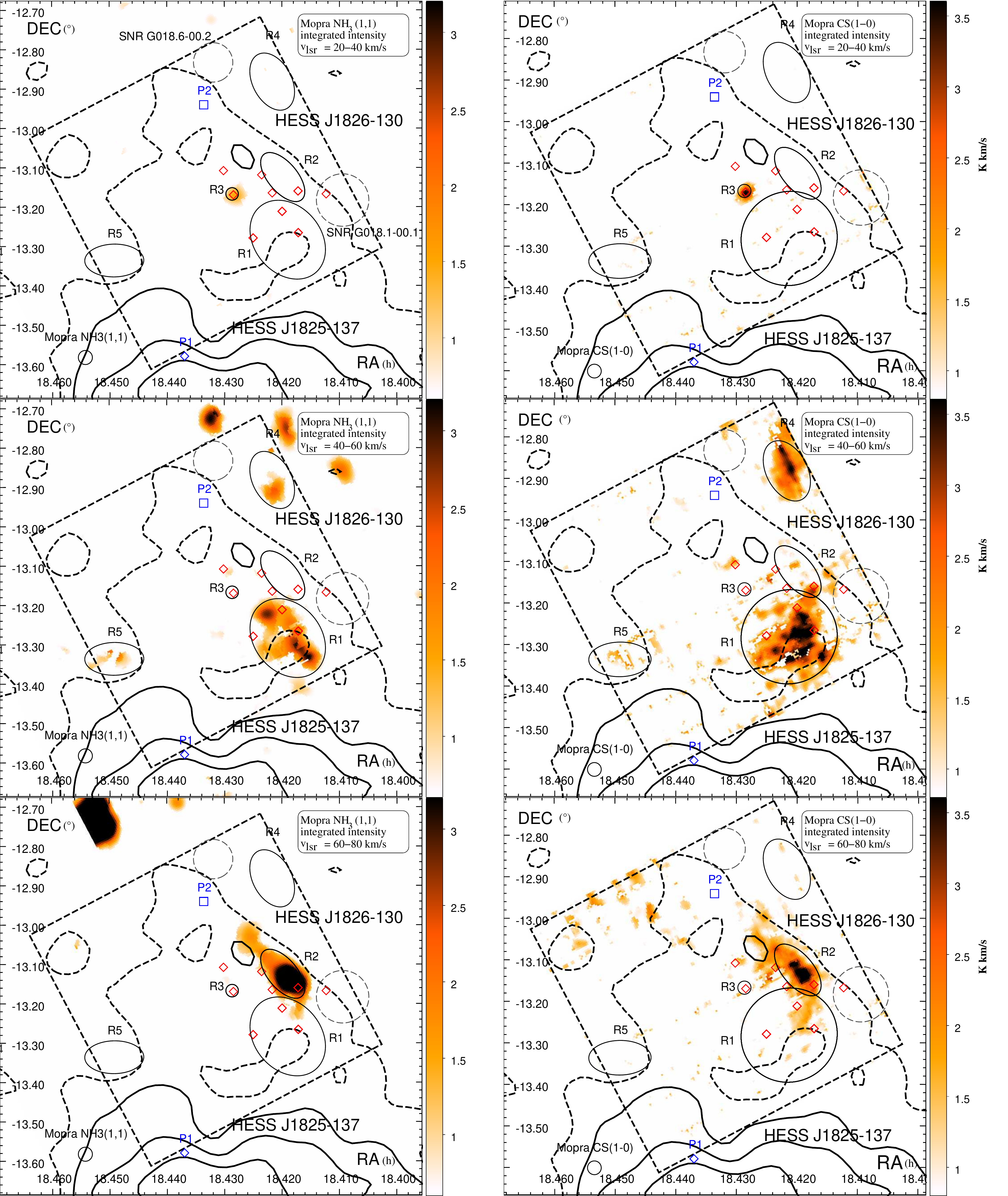}
  \caption{Mopra NH$_{3}$(1,1) and CS(1--0) integrated intensity between  $v_{\text{lsr}}=20-40$\,km/s,  $v_{\text{lsr}}=40-60$\,km/s and  $v_{\text{lsr}}=60-80$\,km/s overlaid by the different regions 
  where NH$_3$(1,1) and CS(1--0) were detected. The region covered by our 7mm survey is shown in black dashed box. The diamonds (red in colour version) indicate the different H\textsc{ii}  regions shown in the SIMBAD database (see \citealt{Anderson2014} for latest H\textsc{ii}  regions catalogue)  while the SNRs are shown in black dashed circles with 
  their labels shown in the top panels. 
  }
 \label{CSCOGRS}
 \end{minipage}
\end{figure*}

\begin{figure}
   \includegraphics[width=0.5\textwidth]{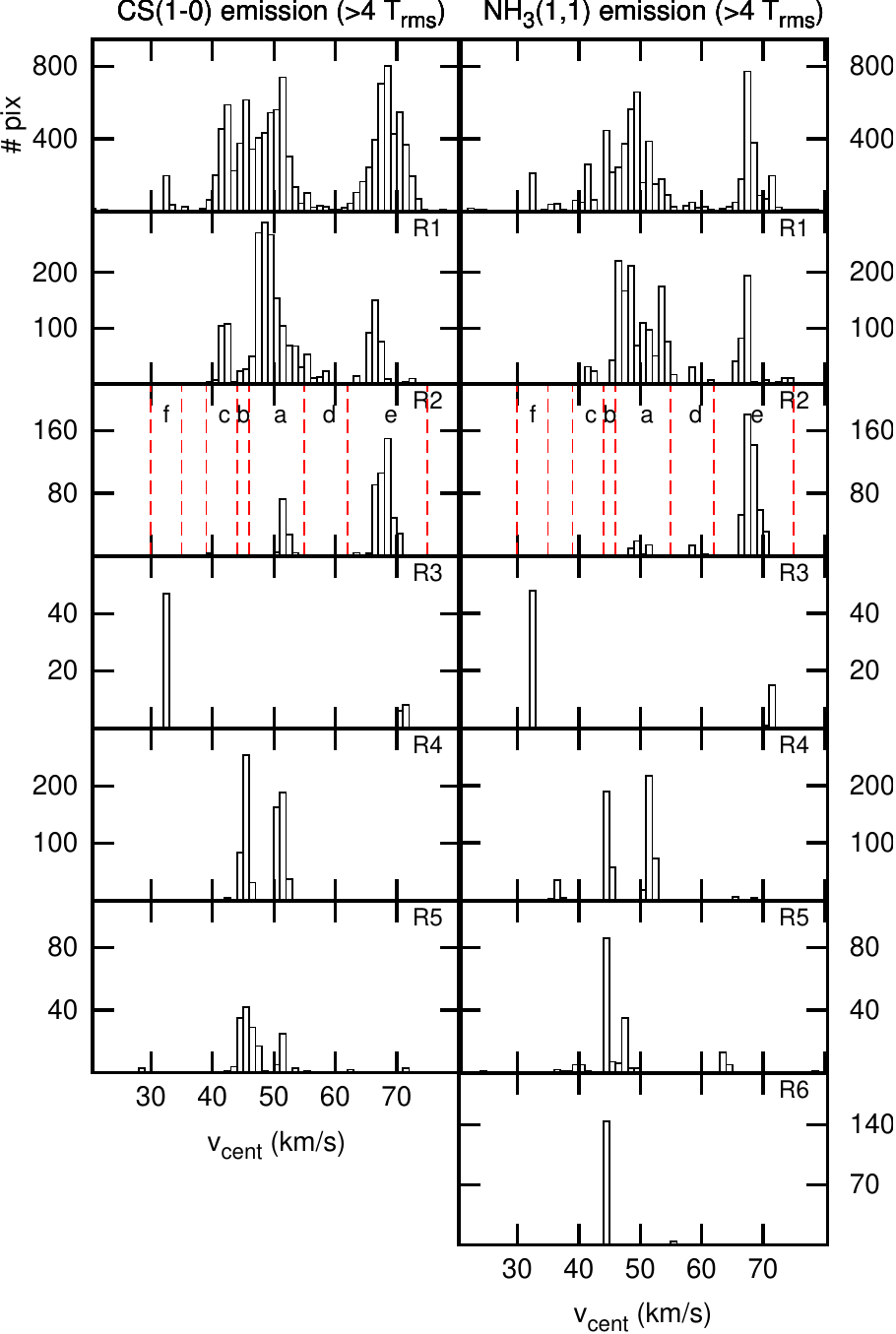}%
   \caption{CS(1--0) \textit{(left)} and NH$_3$(1,1) \textit{(right)} binned centroid velocity distribution (1 km/s intervals) per pixel  of the components detected above
   4 $T_{\text{rms}}$ towards the region covered by our 7mm survey and more specifically the regions \textit{R1} to \textit{R6}. The sub-components `a' to `f' referring to distinct velocity groups (see text and Tables \ref{fitlabel} and \ref{fitlabelbis}) are delimited by the 
   dashed-red vertical lines (see online version).
   }
  \label{NR1vel}
  \end{figure}
  
\begin{figure*}
\begin{minipage}{0.8\textwidth}
 \centering
   \includegraphics[width=\textwidth,angle=0]{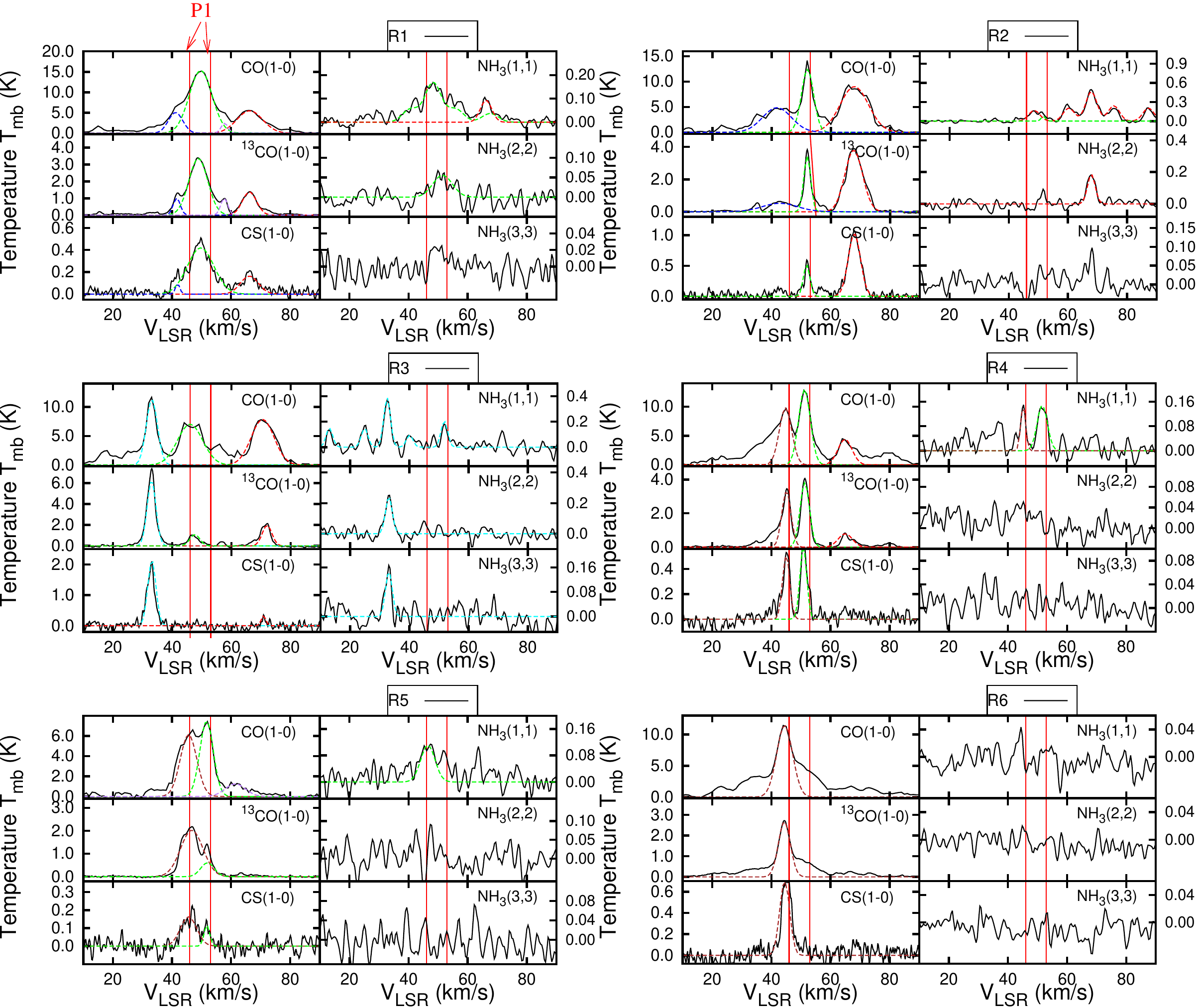}
   \caption{Averaged CO(1--0), $^{13}$CO(1--0), CS(1--0) and NH$_3$(1,1) to (3,3) spectra for the various regions as labelled in Fig.\,\ref{dissociation} and Fig.\,\ref{CSCOGRS}. 
    The vertical lines represent the estimated kinematic velocity range for PSR\,J1826--1334 (P1 in Fig.\,\ref{dissociation}).
    Gaussian fits are represented as dashed lines whose colour code indicates the following velocity $v_{\text{lsr}}$ ranges in km/s ($30-35$, $39-44$, $44-46$, $46-55$, $55-62$, $62-75$) = (cyan, blue, brown, green, purple, red).
    We emphasize that the emission located at $v_{\text{lsr}}=46-55$\,km/s matches the kinematic distance of pulsar P1.}
    \label{SPEC}
    \end{minipage}	
\end{figure*}

\begin{figure*}
 \begin{minipage}{\textwidth}
 \centering
 \hbox{
   \includegraphics[width=0.53\textwidth]{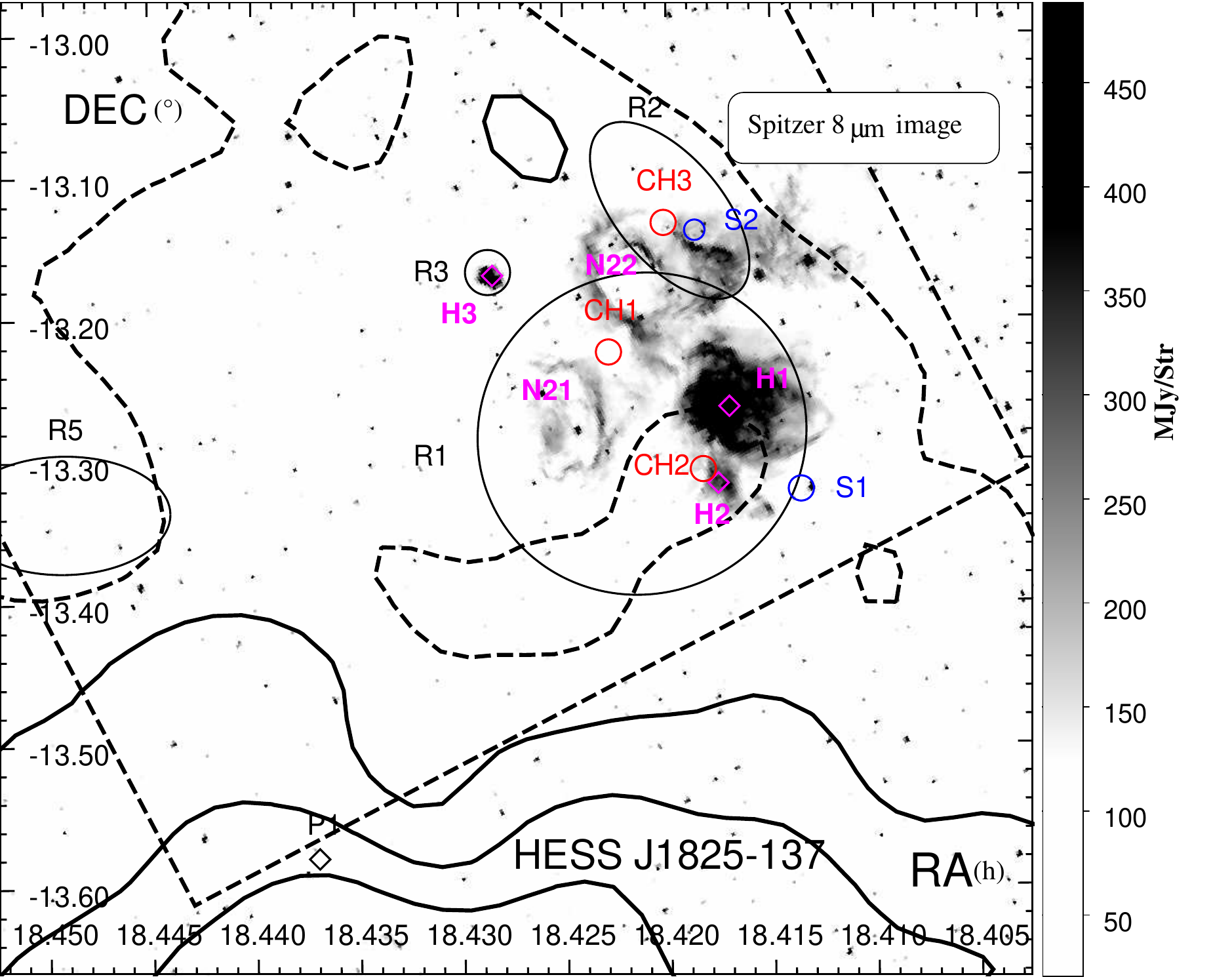}%
   \hspace{-0.2cm}
   \includegraphics[width=0.45\textwidth]{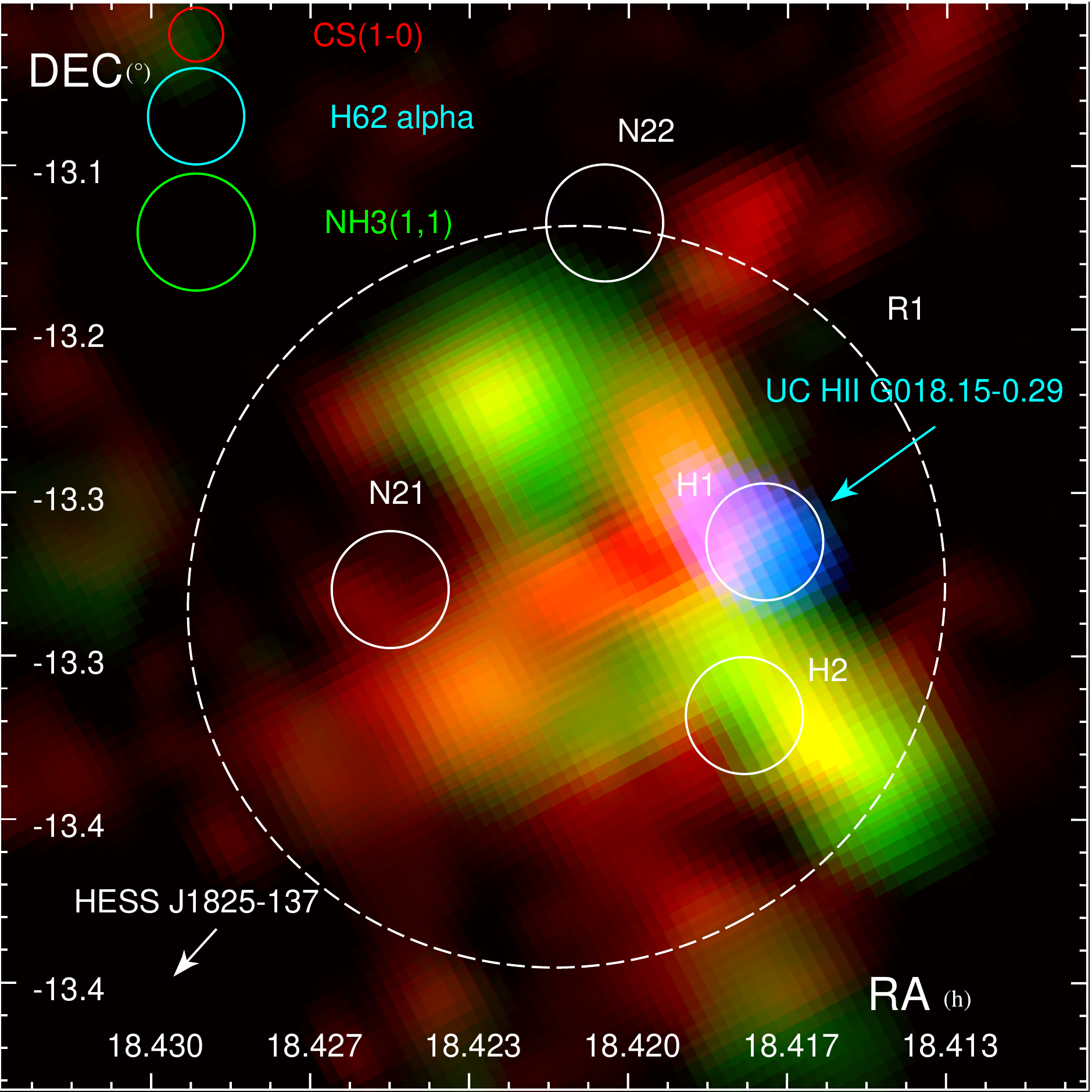}
   }
 \caption{\textit{(Left)} Spitzer 8$\mu$m image towards HESS\,J1826$-$130 (black contours).
 The red circles labelled \textit{CH1} to \textit{CH3} reveal the location of the 44 GHz CH$_3$OH(7$_1$--6$_0$) emission while the blue circles (see online version) \textit{S1} and \textit{S2} indicate SiO(1--0,v=0) and SiO(1--0,v=2) emission.
 \textit{H1} and \textit{H2} represent the ultra-compact H\textsc{ii}  region G018.15-0.29 and the H\textsc{ii}  region G018.142-0.302 respectively  \citep{HDRS2011} while \textit{H3} combines the H\textsc{ii}  regions G018.303-0.389 and G018.305-0.392 \citep{White2005}.
 N21 and N22 shown in purple indicate the location of two IR bubbles \citep{Churchwell2006}. The region covered by our 7mm survey is shown as a black dashed rectangle.
 \textit{(Right)} Three colour image showing the CS(1--0) (red) and NH$_3$ (green) integrated intensity between $v_{\text{lsr}}=40-60$\,km/s and the H62$\alpha$ integrated intensity (blue) between $v_{\text{lsr}}=45-65$\,km/s towards
 \textit{R1a}. The aforementioned H\textsc{ii}  regions are shown as white circles.}%
 \label{CH3spitzer}
\end{minipage}
\end{figure*}

\begin{figure*}
 \begin{minipage}{\textwidth}
 \hbox{
   \includegraphics[width=\textwidth,angle=0]{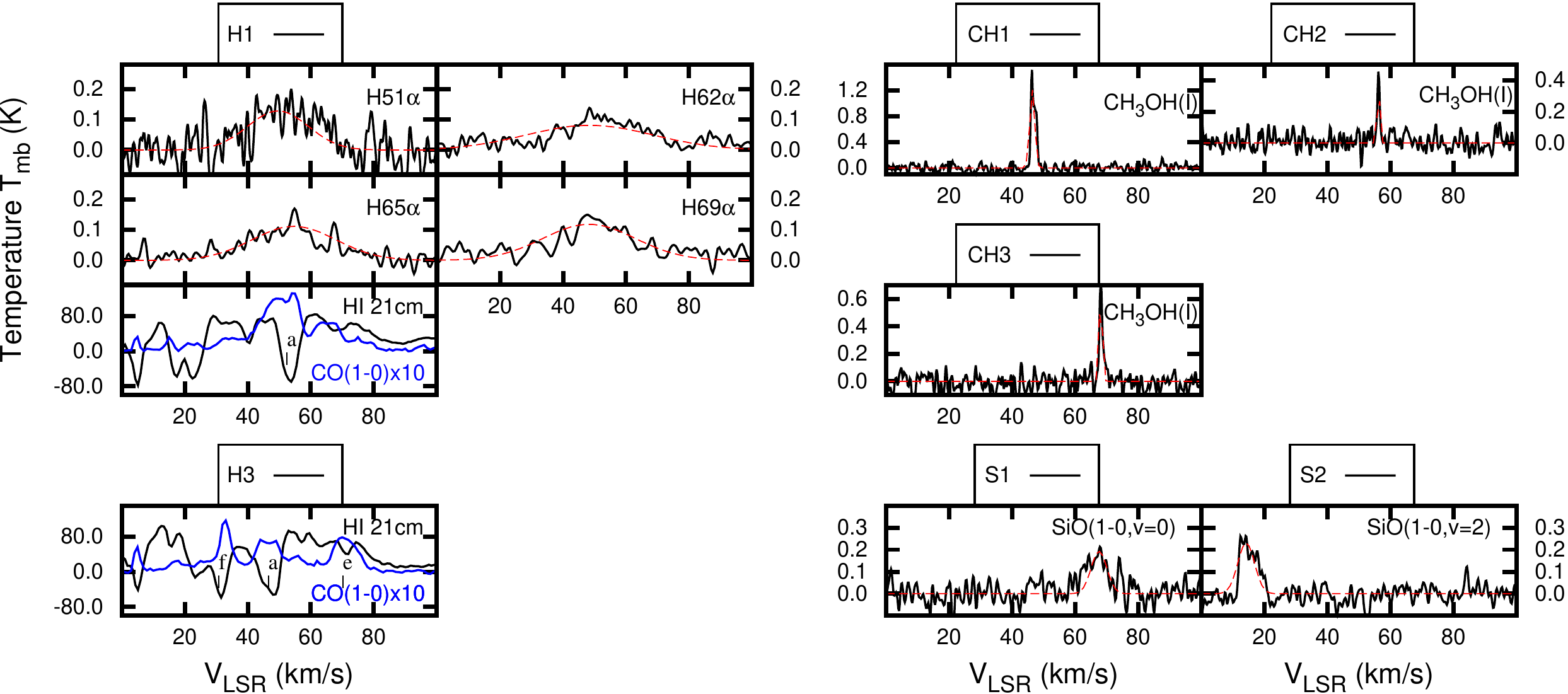}%
   }
 \caption{Spectral profile of the 44GHz maser CH$_3$OH(I), SiO(1--0), CO(1--0), H\textsc{i} 21 cm  and recombination line detections  from the various regions shown in Fig. \ref{CH3spitzer}. The dashed line (red in online version) represents the Gaussian fits to the emission.
 In the CO(1--0) vs H\textsc{i} plots, the absorptions features are indicated by black vertical lines where the labels `a' to `f' indicate the velocity group (see Tables\,\ref{fitlabel} and \ref{fitlabelbis}). }%
\label{HIIspectra}
\end{minipage}
\end{figure*}

\subsection{Overview}
Figs. \ref{dissociation} and \ref{CSCOGRS} show the CO(1-0),$^{13}$CO(1-0), CS(1-0), and NH$_{3}$(1,1) integrated intensity maps over three different velocity spans.

As shown in the top panel of Fig. \ref{NR1vel}, we noticed that the velocity distribution of all detections inside the region covered by our 7mm survey mostly peaked in four velocity regions : \mbox{$v_{\text{lsr}}=40$\,km/s},  
\mbox{$v_{\text{lsr}}=45$ km/s},
\mbox{$v_{\text{lsr}}=50$ km/s} and \mbox{$v_{\text{lsr}}=68$ km/s} and the $^{13}$CO(1--0) emission showed distinct structures at each velocities.
We identified six bright regions, that we labelled \textit{R1} to \textit{R6} (see Figs.\,\ref{dissociation} and \ref{CSCOGRS}). Each region showed bright emission from CO, CS and NH$_3$ listed above (see Fig. \ref{SPEC} for spectral plots), with the exception of \textit{R6} whose
NH$_3$(1,1) averaged emission over the region was too weak. The composition of these regions will be discussed in detail in order to provide a better understanding about the complex morphology
of the observed emission.

Fig. \ref{CH3spitzer} shows the 8$\mu$m continuum image from the Spitzer GLIMPSE survey \citep{Spitzer}. We observed  various infra-red (IR) features spatially coincident with  dense molecular gas from some the aforementioned regions.
\citet{Anderson2014} indicated that these IR sources were mostly H\textsc{ii}  regions (see red diamonds in Fig. \ref{CSCOGRS}). We detected several star-forming region tracers that were spatially coincident with these IR sources (which will be detailed 
in the next section). Association between these IR regions and the aforementioned molecular gas regions indicate likely source for driving the gas motion.

\subsection{CO(1--0) and $^{13}$CO(1--0) analysis} 
The CO(1--0) and its isotopologue $^{13}$CO(1--0) averaged emission in the selected regions \textit{R1} to \textit{R6} were fitted with a single Gaussian. 
The components from each region were labelled from `a' to `f' according to the velocity group they belonged to.
For example, `a' represents the velocity range v$_{\text{lsr}}=46-55$\,km/s matching the pulsar P1's kinematic distance, whilst
`b', `c', `d', `e' and `f' indicate the range v$_{\text{lsr}}=44-46$\,km/s, $39-44$\,km/s, $55-62$\,km/s, $62-75$\,km/s and $30-35$\,km/s respectively.
The total mass was also determined using the CO(1--0) averaged integrated intensity $W_{\text{CO}}$ and the conversion factor $X_{\text{CO}}=2.0\times10^{20}$ cm$^{-2}$/(K km/s)  
The conversion factor is generally assumed to be constant across the Galactic plane although its value may slightly vary as a function of the galactocentric radius \citep{Strong}.
Finally, we used a prolate geometry to provide H$_2$ density, $n_{\text{H}_2}$, estimates via Eq. \ref{H2dens}. The total proton density $n_{\text{H}}$ can be deduced using
\mbox{$n_\text{H}$=2.8\,$n_{\text{H}_2}$} which accounts for 20\% He fraction.
We also used the full width half maximum (FWHM) of the isotopologue $^{13}$CO(1--0), less prone to optical depth effects (e.g broadening), to obtain the Virial mass $M_{\text{vir}}$ of the selected regions. We 
used the inverse-squared $r^{-2}$ and Gaussian density distribution (see \citealt{Protheroe2008}) as lower and upper-range of the Virial masses respectively.

Among the observed clouds, the one located at \mbox{$v_{\text{lsr}}=45-60$\,km/s} is at a  kinematic distance $d=4$\,kpc which is similar to that of the pulsar PSR\,J1826$-$1334, is adjacent to HESS\,J1825$-$137
 (see red dashed circle in the middle right panel of Fig.\,\ref{dissociation}).
Assuming the molecular cloud to be spherical with radius $R_{\text{MC}}\sim 18$\,pc and centred at (RA,Dec)=(18.431h,$-13.26^{\circ}$), we obtain from our CO(1--0) observations an averaged density over the region
\mbox{$n_{\text{H}}\sim6.1\times10^{2}$ cm$^{-3}$} and a total mass \mbox{$M_{\text{H}_2}\sim3.3\times10^5\solmass$}.
The density distribution is not uniform  as shown by the presence of several molecular clumps (e.g as in sub-regions labeled \textit{R1} to \textit{R5}) seen in CO(1--0) and CS(1--0).

We also detected via the CO(1--0) and $^{13}$CO(1--0) molecular transitions a diffuse molecular cloud at \mbox{$v_{\text{lsr}}=18$\,km/s} as shown in Fig. \ref{NantenCO1025}. Its kinematic distance  \mbox{$d=1.7$\,kpc} is similar to the pulsar 
PSR\,J1826$-$1256's estimated distance and it sits between the pulsar and the SNR\,G018.6$-$0.2.
 We derived a total mass M$_{\text{H}_2}=5.3\times10^{3} \solmass$ and an averaged density over the region \mbox{$n_{\text{H}}=5.9\times10^{2}$\,cm$^{-3}$} (within the red circle in Fig.\,\ref{NantenCO1025}).
At \mbox{$v_{\text{lsr}}=60-80$\,km/s} (see Fig.\,\ref{dissociation} bottom panel), we also observed that the molecular gas appears adjacent to the pulsar P2 and the \mbox{SNR\,G018.6--0.2} with the bulk of the molecular gas
 located north of the pulsar P2.

\subsection{CS analysis}

From the different spectra shown in Fig. \ref{SPEC}, CS(1--0) components were also fitted with a single Gaussian and the fit parameters have been listed in Table \ref{fitlabel} and Table \ref{fitlabelbis}.
In order to derive physical parameters of the different regions, we used the local thermal equilibrium assumption (LTE).
In the case where the isotopologue C$^{34}$S(1--0) were also detected, we estimated of the averaged optical depth  $\tau_{\text{CS(1-0)}}$ using Eq. \ref{tauCS}. An optically thin 
scenario \mbox{$\tau_{\text{CS(1-0)}}$=0} would otherwise be used to obtain the column density of the upper state $N_{\text{CS}_1}$ via Eq. \ref{CSeq2}. The averaged C$^{34}$S(1--0) spectra detected in some 
of the studied regions \textit{R1},\textit{R2},\textit{R3} are shown in Fig.\,\ref{34CSregions}.

We used the estimated kinetic temperature $T_{\text{kin}}$ from our NH$_3$ analysis (see below) to obtain the total column density $N_{\text{CS}}$ using Eq. \ref{CSeq3}. This assumption was only valid
if our NH$_3$ and CS tracers probed the same gas. In all other cases, we assumed $T_{\text{kin}}$=10 K.

In order to obtain the H$_2$ column density $N_{\text{H}_2}$, we chose the abundance ratio \mbox{$\chi_{\text{CS}}=4\times10^{-9}$} which was in the range of values \mbox{$\chi_{\text{CS}}=10^{-9}\rightarrow10^{-8}$} indicated by \citet{Irvine1987} who studied the chemical abundances inside 
several distinct regions e.g Orion KL and Sgr B2. \citet{Zitchenko1994} also chose this abundance ratio for the study of several CS cores. 
The derived  H$_2$ parameters are scaled by the abundance ratio and thus may vary by a factor of two.

We provided total mass estimates by using the kinematic distance in Eq. \ref{H2mass} and considering the molecular gas consisted of 20\% Helium.
As per our CO analysis, we used a prolate geometry to provide H$_2$ density $n_{\text{H}_2}$ estimates.

\subsection{NH$_{3}$ analysis}
To fit the emission of the NH$_3$(1,1) inversion transition, we used five Gaussians separated by known velocities to fit the main peak and the four satellite lines \citep{Wilson}. The fit parameters of each regions were listed in Table \ref{fitlabelbis}.
 
Whenever we detected NH$_3$(1,1) satellite lines, we used the ratio of the integrated intensity between the main and satellite line and used Eq. \ref{tauNH3} to determine the averaged main line 
optical depth. Finally, based on the NH$_3$(1,1) partition function, we used Eq. \ref{tauNH3b} to estimate the optical depth of the NH$_3$(1,1) emission $\tau_{\text{NH}_3\text{(1,1)}}$.

As per our CS analysis, we approximated our regions to be in LTE and used Eq. \ref{NH3eq4} to obtain the NH$_3$(1,1) column density. 
If NH$_3$(2,2) emission was also detected, we obtained the temperature $T_{\text{kin}}$ using Eqs. \ref{NH3eq6} and \ref{NH3eq7}. This method remains only valid for kinetic temperature below 40K.
We then considered an even chemical abundance between ortho-NH$_3$ and para-NH$_3$ to obtain the total column density $N_{\text{NH}_3}$ via Eq. \ref{NH3eq5}.
To convert the NH$_3$ column density into H$_2$ column density, we used an abundance ratio \mbox{$\chi_{\text{NH}_3}=1\times 10^{-8}$} which is in the range provided by \citet{Irvine1987}.
Finally, the same method as per our CS and CO analysis was used to determine the mass and density estimates.
From Fig.\,\ref{CSCOGRS}, we find that the morphology of the gas detected by the NH$_3$(1,1) inversion transition coincides with the CS(1--0) emission towards the region covered by our 7mm survey.

\subsection{HI analysis}
In order to complete the picture of the gas distribution towards HESS\,J1826$-$130, we made use of SGPS data with $\Delta v=0.8$\,km/s and $T_{\text{rms}}$=1.4\,K/channel \citep{SGPS2005} to
 search for diffuse atomic gas.
 
 Comparing H\textsc{i} and CO(1--0) is  an effective method to provide kinematic distance ambiguity resolution of molecular clouds (KDAR, see \citealt{Anderson,Roman} for further details) provided we know the location
 of the continuum source appearing in the line of sight.
 For instance, Fig.\,\ref{HIIspectra} shows the association between the CO(1--0) emission (blue lines) and the H\textsc{i} absorption (black lines) in region \textit{H1} and \textit{H3}.
 
\begin{figure}
 \includegraphics[width=0.5\textwidth]{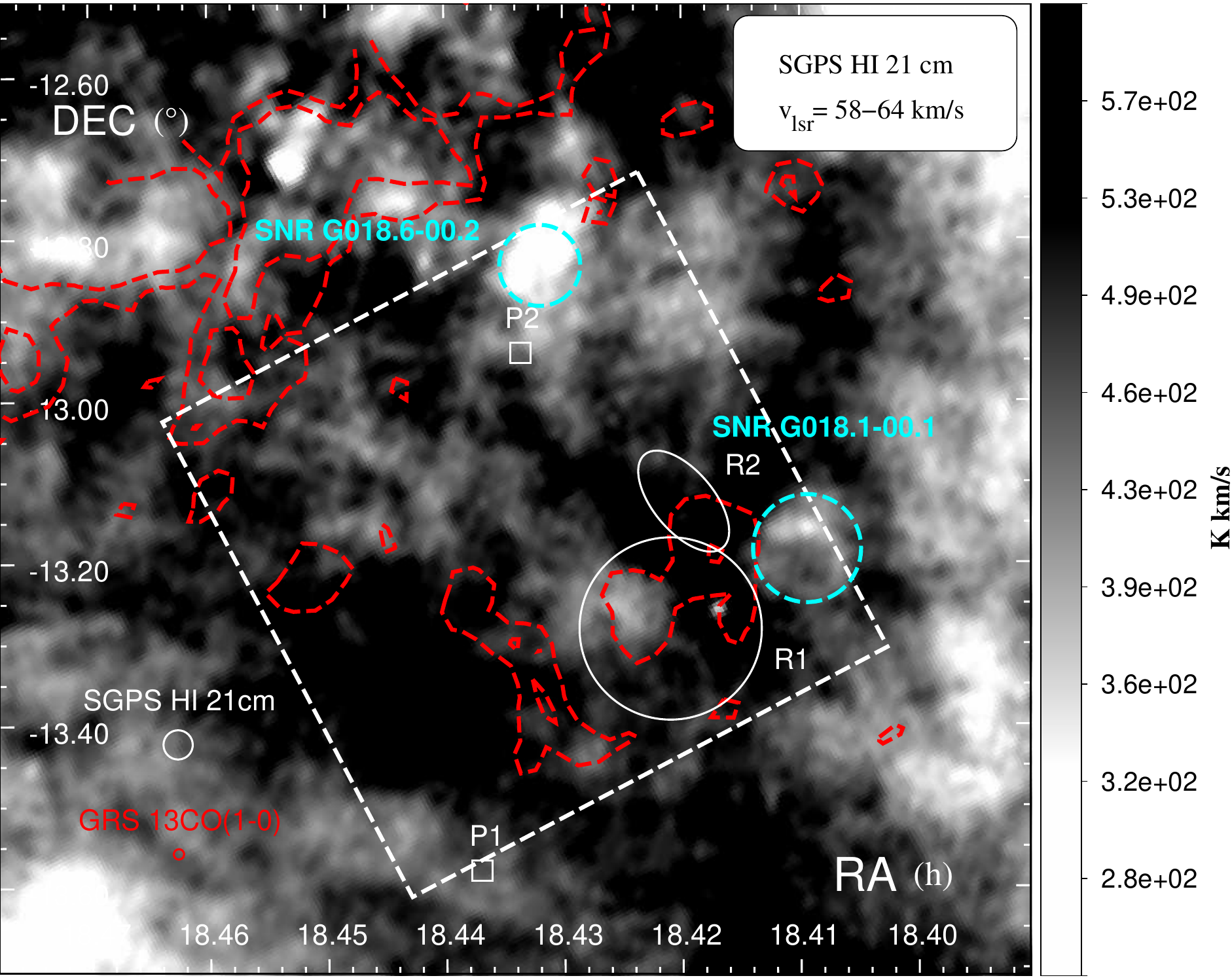}
 \caption{H\textsc{i} 21 cm integrated intensity in grey-scale between $v_{\text{lsr}}=58-64$\,km/s overlaid by the GRS $^{13}$CO(1--0) contours in red (2.0 and 3.0 K km/s, see colour version). The white ellipses indicate the position of \textit{R1}  and \textit{R2},
  while cyan circles show the two SNRs. The white squares represent the pulsars PSR\,J1826$-$1334 (P1) and PSR\,J1826$-$1256 (P2). The white dashed box represent our Mopra 7mm coverage.}
 \label{HIsnapshot}
\end{figure}

 We also use the H\textsc{i} data to probe potential dips associated with energetic sources (e.g SNRs) in order to provide an estimate of their kinematic distance.
 Fig.\,\ref{HIsnapshot} shows the H\textsc{i} integrated intensity towards HESS\,J1826$-$130 between \mbox{$v_{\text{lsr}}=58-64$\,km/s}.
 We noticed a dip in H\textsc{i} emission towards SNR\,G018.6$-$0.2 which did not overlap with the $^{13}$CO(1--0) contour shown in red. 
 From the GRS $^{13}$CO(1--0) longitude-velocity plot in Fig.\,\ref{GRSPVplot}, we also noted a lack of emission at \mbox{$v_{\text{lsr}}=60-70$\,km/s} spatially coincident with this SNR position, with weak emission at $v_{\text{lsr}}\sim60$\,km/s and $\sim75$\,km/s.
 These weak $^{13}$CO(1--0) features may provide further evidence of a shell towards this SNR where molecular gas have been accelerated.
 Thus, we argue a shell has been produced by SNR\,G018.6$-$0.2's  progenitor star located at \mbox{$v_{\text{lsr}}\sim60-64$\,km/s} inferring a SNR distance \mbox{$d=4.5$\,kpc} (near) or \mbox{$11.4$\,kpc} (far).

\subsection{Discussion of individual regions}

\begin{table*}
\caption{Derived parameters of the Gaussian fits from the selected regions in Fig. \ref{dissociation} (see text). $v_{\text{cent}}$ represents the velocity centroid of the Gaussian while $\Delta$v indicates the Gaussian FWHM. W=$\int{T_{mb}\text{d}v}$ represents the integrated main beam intensity applying the main beam correction factor $\eta_{\text{mb}}$ (see text for details).
   The components detected by the different tracers CS(1--0), $^{13}$CO(1-0), CO(1-0) are labeled as `a,b,c,d,e,f' according to their velocity range (see footnote below table).
   } 
   \label{fitlabel}
 \begin{minipage}{\textwidth}
\centering
  \begin{tabular}{lllllllllll}
    \toprule
   CO(1--0)& \multicolumn{4}{l}{R1}& \multicolumn{3}{l}{R2}& \multicolumn{3}{l}{R3}\\
   & \multicolumn{4}{l}{(RA,Dec)=(18.421h, $-13.282^{\circ}$)}&\multicolumn{3}{l}{(RA,Dec)=(18.420h, $-13.125^{\circ}$)}&\multicolumn{3}{l}{(RA,Dec)=(18.429h, $-13.178^{\circ}$)}\\
   & \multicolumn{4}{l}{Radii\textsuperscript{*}=($405^{\prime\prime}\times405^{\prime\prime}$)}& \multicolumn{3}{l}{Radii=($135^{\prime\prime}\times270^{\prime\prime}$)}&\multicolumn{3}{l}{Radii=($64^{\prime\prime}\times64^{\prime\prime}$)}  \\
          &a$ $&c$ $&d$ $&e$ $&a$ $&c$ $&e$ $&a$ $&e$ $&f$ $ \\
   \midrule
   Peak value $T_{\text{A}}^{*}$ (K)&13.6&4.5&2.7&4.9&11.0&4.3&8.0&6.2&7.0&10.0 \\
   $v_{\text{cent}}$ (km/s)&49.6&41.2&57.8&66.2&52.0&41.8&67.9&46.0&70.5&33.1\\
   $\Delta v$ (km/s)&9.0&5.7&2.8&10.3&4.8&11.4&11.3&10.0&9.1&4.5 \\
   $W_{\text{CO}}$ (K km/s)&148.2&31.0&7.2&60.0&64.4&59.2&108.3&75.6&76.6&53.6 \\
   $T_{\text{rms}}$/ch/$\sqrt{\text{bins}}$ (K)& \multicolumn{4}{l}{---0.07---}& \multicolumn{3}{l}{---0.18---} & \multicolumn{1}{l}{---0.30---}  \\
   \\
    \toprule
   $^{13}$CO(1--0)&\multicolumn{4}{l}{R1}& \multicolumn{3}{l}{R2}& \multicolumn{3}{l}{R3}\\
     &a&c&d&e&a&c&e&a&e&f \\
   \midrule
   Peak value $T_{\text{A}}^{*}$ (K)&1.6&0.5&0.5&0.7&1.7&0.3&1.8&0.5&0.9&3.0 \\
   $v_{\text{cent}}$ (km/s)&48.9&41.6&57.7&66.2&52.0&42.8&67.7&47.6&72.0&33.0\\
   $\Delta v$ (km/s)&7.5&2.8&2.8&5.6&2.8&13.9&7.1&3.9&3.9&3.4 \\
   $W_{^{13}\text{CO}}$ (K km/s)&27.0&3.1&7.2&8.3&10.1&7.7&28.5&4.6&8.1&23.2 \\
   $T_{\text{rms}}$/ch/$\sqrt{\text{bins}}$ (K)& \multicolumn{4}{l}{---0.01---}& \multicolumn{3}{l}{---0.01---} & \multicolumn{1}{l}{---0.03---}  \\
   \\
   \toprule
   CS(1--0)&\multicolumn{4}{l}{R1}& \multicolumn{3}{l}{R2}& \multicolumn{3}{l}{R3}\\
   &a&c&d&e&a&c&e&a&e&f \\
   \midrule
   Peak value $T_{\text{A}}^{*}$ (K)&0.2&$<0.1$&-&0.1&0.2&-&0.4&-&0.1&0.9 \\
   $v_{\text{cent}}$ (km/s)&49.7&41.9&-&66.2&51.8&-&67.9&-&70.8&33.1\\
   $\Delta v$ (km/s)&10.7&1.9&-&7.2&2.6&-&4.8&-&1.8&3.2 \\
   $W_{\text{CS}}$ (K km/s)&4.7&0.2&-&1.3&1.4&-&5.3&-&0.7&7.5 \\
   $T_{\text{rms}}$/ch/$\sqrt{\text{bins}}$ (K)& \multicolumn{4}{l}{---0.01---}& \multicolumn{3}{l}{---0.01---} & \multicolumn{3}{l}{---0.03---}  \\
   \bottomrule
   \end{tabular}
   \vspace{0.5cm}
   \begin{tabular}{llllllll}
    \toprule
   CO(1--0)& \multicolumn{3}{l}{R4}& \multicolumn{3}{l}{R5}& \multicolumn{1}{l}{R6}\\
   & \multicolumn{3}{l}{(RA,Dec)=(18.422h, $-12.832^{\circ}$)}&\multicolumn{3}{l}{(RA,Dec)=(18.449h, $-13.336^{\circ}$)}&\multicolumn{1}{l}{(RA,Dec)=(18.385h, $-14.049^{\circ}$)}\\
   & \multicolumn{3}{l}{Radii=($175^{\prime\prime}\times280^{\prime\prime}$)}& \multicolumn{3}{l}{Radii=($150^{\prime\prime}\times270^{\prime\prime}$)}&\multicolumn{1}{l}{Radii=($460^{\prime\prime}\times460^{\prime\prime}$)}  \\
          &a$ $&b$ $&e$ $&a$ $&b$ $&d$ $& b\\
   \midrule
   Peak value $T_{\text{A}}^{*}$ (K)&11.5&8.7&3.9&6.6&5.4&1.1&10.2 \\
   $v_{\text{cent}}$ (km/s)&51.2&44.8&64.9&51.7&45.6&61.4&44.5\\
   $\Delta v$ (km/s)&4.6&4.9&5.6&5.4&6.5&7.9&6.3\\
   $W_{\text{CO}}$ (K km/s)&64.5&51.5&26.3&43.0&43.0&10.6&77.1 \\
   $T_{\text{rms}}$/ch/$\sqrt{\text{bins}}$ (K)& \multicolumn{3}{l}{---0.23---}& \multicolumn{3}{l}{---0.24---} & \multicolumn{1}{l}{---0.10---}  \\
   \\
    \toprule
   $^{13}$CO(1--0)&\multicolumn{3}{l}{R4}& \multicolumn{3}{l}{R5}& \multicolumn{1}{l}{R6}\\
     &a&b&e&a&b&d&b \\
   \midrule
   Peak value $T_{\text{A}}^{*}$ (K)&1.8&1.6&0.4&0.3&1.0&-&1.3\\
   $v_{\text{cent}}$ (km/s)&51.3&45.1&65.0&52.0&46.7&-&44.4\\
   $\Delta v$ (km/s)&3.7&3.6&5.6&5.4&8.9&-4.8\\
   $W_{^{13}\text{CO}}$ (K km/s)&15.3&13.0&4.7&3.5&19.2&-&13.6 \\
   $T_{\text{rms}}$/ch/$\sqrt{\text{bins}}$ (K)& \multicolumn{3}{l}{---0.03---}& \multicolumn{3}{l}{---0.01---} & \multicolumn{1}{l}{---0.01---}  \\
   \\
   \toprule
   CS(1--0)&\multicolumn{3}{l}{R4}& \multicolumn{3}{l}{R5}& \multicolumn{1}{l}{R6}\\
   &a&b&e&a&b&d&b \\
   \midrule
   Peak value $T_{\text{A}}^{*}$ (K)&0.3&0.2&-&0.1&0.1&-&0.3 \\
   $v_{\text{cent}}$ (km/s)&50.8&45.0&-&51.8&46.7&-&44.5\\
   $\Delta v$ (km/s)&2.6&2.8&-&2.2&4.7&-&4.0 \\
   $W_{\text{CS}}$ (K km/s)&1.6&1.6&-&0.3&0.9&-&3.0\\
   $T_{\text{rms}}$/ch/$\sqrt{\text{bins}}$ (K)& \multicolumn{3}{l}{---0.01---}& \multicolumn{3}{l}{---0.01---} & \multicolumn{1}{l}{---0.01---}  \\
   \bottomrule
   \multicolumn{8}{l}{\footnotesize{ $^{*}$ Radii represents the dimensions of the ellipse (semi-minor axis $\times$ semi-major axis).}}\\
   \multicolumn{8}{l}{\footnotesize{component a : $v_{\text{lsr}}=46-55$ km/s, matching the dispersion measure of P1. }}\\
   \multicolumn{8}{l}{\footnotesize{component b : $v_{\text{lsr}}=44-46$ km/s.}}\\
   \multicolumn{8}{l}{\footnotesize{component c : $v_{\text{lsr}}=39-44$ km/s.}}\\
   \multicolumn{8}{l}{\footnotesize{component d : $v_{\text{lsr}}=55-62$ km/s.}}\\
   \multicolumn{8}{l}{\footnotesize{component e : $v_{\text{lsr}}=62-75$ km/s.}}\\
   \multicolumn{8}{l}{\footnotesize{component f : $v_{\text{lsr}}=30-35$ km/s.}}
   \end{tabular}

   \end{minipage}
   \end{table*}

   \begin{table*}
 \begin{minipage}{\textwidth}
 \caption{Derived parameters of the five Gaussian fits used to model NH$_3$(1,1) emission, and the one Gaussian fit for the NH$_3$(2,2) emission averaged over the region shown in Fig. \ref{CSCOGRS}.
 $T_{\text{A}_m}^{*}$ indicates the peak intensity of the main emission while $T_{\text{A}_s}^{*}$ are the peak intensities of the surrounding satellite lines. $v_{\text{cent}}$ represents the velocity centroid of the Gaussian while $\Delta$v indicates the Gaussian FWHM. Finally W=$\int{T_{mb}\text{d}v}$ represents the integrated main beam intensity applying the main beam correction factor $\eta_{\text{mb}}$ (see text for details).
   The different components detected by the different tracers CS(1-0), $^{13}$CO(1-0), CO(1-0) are labeled as `a,b,c,d,e,f' according to their velocity range (see footnote below table). }
  \label{fitlabelbis} 
\centering
  \begin{tabular}{lllllllllll}
    \toprule
   NH$_3$(1,1)& \multicolumn{4}{l}{R1}& \multicolumn{3}{l}{R2}& \multicolumn{3}{l}{R3}\\
   & \multicolumn{4}{l}{(RA,Dec)=(18.419h, $-13.284^{\circ}$)}&\multicolumn{3}{l}{(RA,Dec)=(18.420h, $-13.125^{\circ}$)}&\multicolumn{3}{l}{(RA,Dec)=(18.429h, $-13.178^{\circ}$)}\\
   & \multicolumn{4}{l}{Radii=($395^{\prime\prime}\times305^{\prime\prime}$)}& \multicolumn{3}{l}{Radii=($135^{\prime\prime}\times270^{\prime\prime}$)}&\multicolumn{3}{l}{Radii=($64^{\prime\prime}\times64^{\prime\prime}$)}  \\
          &a$ $&c$ $&d$ $&e$ $&a$ $&c$ $&e$ $&a$ $&e$ $&f$ $ \\
   \midrule
   Peak value $T_{\text{A}_m}^{*}$ (K)&0.09&-&-&0.05&0.04&-&0.25&-&-&0.21 \\
   Peak value $T_{\text{A}_{s1}}^{*}$ (K)&0.03&-&-&-&-&-&0.13&-&-&0.05 \\
   Peak value $T_{\text{A}_{s2}}^{*}$ (K)&0.05&-&-&-&-&-&0.11&-&-&0.05 \\
   Peak value $T_{\text{A}_{s3}}^{*}$ (K)&0.02&-&-&-&-&-&0.11&-&-&0.08 \\
   Peak value $T_{\text{A}_{s4}}^{*}$ (K)&-&-&-&-&-&-&0.09&-&-&0.08\\
   $v_{\text{cent}}$ (km/s)&48.3&-&-&66.2&51.7&-&67.9&-&-&33.1\\
   $\Delta v$ (km/s)&6.4&-&-&4.5&1.8&-&4.5&-&-&2.7\\
   $W_{\text{NH}_3\text(1,1)}$ (K km/s)&2.2&-&-&0.8&1.4&-&5.3&-&-&7.5 \\
   $T_{\text{rms}}$/ch/$\sqrt{\text{bins}}$ (K)& \multicolumn{4}{l}{---0.01---}& \multicolumn{3}{l}{---0.01---} & \multicolumn{3}{l}{---0.01---}  \\
   \\
    \toprule
     NH$_3$(2,2)& \multicolumn{4}{l}{R1}& \multicolumn{3}{l}{R2}& \multicolumn{3}{l}{R3}\\
     &a$ $&c$ $&d$ $&e$ $&a$ $&c$ $&e$ $&a$ $&e$ $&f$ $ \\
   \midrule
   Peak value $T_{\text{A}}^{*}$ (K)&0.03&-&-&-&-&-&0.10&-&-&0.13\\
   $v_{\text{cent}}$ (km/s)&51.5&-&-&-&-&-&67.9&-&-&33.1\\
   $\Delta v$ (km/s)&7.7&-&-&-&-&-&3.1&-&-&2.7 \\
   $W_{\text{NH}_3\text(2,2)}$ (K km/s)&0.5&-&-&-&-&-&0.6&-&-&0.7 \\
   $T_{\text{rms}}$/ch/$\sqrt{\text{bins}}$ (K)& \multicolumn{4}{l}{---0.01---}& \multicolumn{3}{l}{---0.01---} & \multicolumn{3}{l}{---0.02---}  \\
   \\
    \toprule
    \end{tabular}
    \vspace{0.5cm}
    
    \begin{tabular}{llllllll}
    \toprule
   NH$_3$(1,1)& \multicolumn{3}{l}{R4}& \multicolumn{3}{l}{R5}& \multicolumn{1}{l}{R6}\\
   & \multicolumn{3}{l}{(RA,Dec)=(18.422h, $-12.832^{\circ}$)}&\multicolumn{3}{l}{(RA,Dec)=(18.449h, $-13.336^{\circ}$)}&\multicolumn{1}{l}{(RA,Dec)=(18.385h, $-14.049^{\circ}$)}\\
   & \multicolumn{3}{l}{Radii=($175^{\prime\prime}\times280^{\prime\prime}$)}& \multicolumn{3}{l}{Radii=($150^{\prime\prime}\times270^{\prime\prime}$)}&\multicolumn{1}{l}{Radii=($460^{\prime\prime}\times460^{\prime\prime}$)}  \\
          &a$ $&b$ $&e$ $&a$ $&b$ $&d$ $&b\\
   \midrule
   Peak value $T_{\text{A}_m}^{*}$ (K)&0.08&0.08&-&0.06&-&-&- \\
   Peak value $T_{\text{A}_{s1}}^{*}$ (K)&-&-&-&-&-&-&- \\
   Peak value $T_{\text{A}_{s2}}^{*}$ (K)&-&-&-&-&-&-&-\\
   Peak value $T_{\text{A}_{s3}}^{*}$ (K)&-&-&-&-&-&-&- \\
   Peak value $T_{\text{A}_{s4}}^{*}$ (K)&-&-&-&-&-&-&- \\
   $v_{\text{cent}}$ (km/s)&45.1&51.6&-&46.5&-&-&-\\
   $\Delta v$ (km/s)&2.0&4.0&-&5.6&-&-&- \\
   $W_{\text{NH}_3\text{(1,1)}}$ (K km/s)&0.3&0.6&-&0.7&-&-&- \\
   $T_{\text{rms}}$/ch/$\sqrt{\text{bins}}$ (K)& \multicolumn{3}{l}{---0.01---}& \multicolumn{3}{l}{---0.01---} & \multicolumn{1}{l}{---0.01---}  \\
   \\
    \toprule
    NH$_3$(2,2)& \multicolumn{3}{l}{R4}& \multicolumn{3}{l}{R5}& \multicolumn{1}{l}{R6}\\
     &a$ $&b$ $&c$ $&a$ $&b$ $&c$ $& \\
   \midrule
   Peak value $T_{\text{A}}^{*}$ (K)&-&-&-&-&-&-&- \\
   $v_{\text{cent}}$ (km/s)&-&-&-&-&-&-&-\\
   $\Delta v$ (km/s)&-&-&-&-&-&-&- \\
   $W_{\text{NH}_3\text(2,2)}$ (K km/s)&-&-&-&-&-&-&- \\
   $T_{\text{rms}}$/ch/$\sqrt{\text{bins}}$ (K)& \multicolumn{3}{l}{---0.01---}& \multicolumn{3}{l}{---0.01---} & \multicolumn{1}{l}{---0.02---}  \\
   \\
    \toprule
    \multicolumn{8}{l}{\footnotesize{ $^{*}$ Radii represents the dimensions of the ellipse (semi-minor axis $\times$ semi-major axis).}}\\
   \multicolumn{8}{l}{\footnotesize{component a : $v_{\text{lsr}}=46-55$ km/s, matching the dispersion measure of P1.}}\\
   \multicolumn{8}{l}{\footnotesize{component b : $v_{\text{lsr}}=44-46$ km/s.}}\\
   \multicolumn{8}{l}{\footnotesize{component c : $v_{\text{lsr}}=39-44$ km/s.}}\\
   \multicolumn{8}{l}{\footnotesize{component d : $v_{\text{lsr}}=55-62$ km/s.}}\\
   \multicolumn{8}{l}{\footnotesize{component e : $v_{\text{lsr}}=62-75$ km/s.}}\\
   \multicolumn{8}{l}{\footnotesize{component f : $v_{\text{lsr}}=30-35$ km/s.}}
    \end{tabular}
    \end{minipage}
    \end{table*}

\subsubsection{Region R1}
\label{section5}

Region \textit{R1} (RA=18.421h, Dec=$-13.28^{\circ}$) is located 0.5$^{\circ}$ away from the pulsar PSR\,J1826-1334 and contains the bulk of the $^{13}$CO(1--0) and CS(1--0) emission. 
From the CO(1-0) and $^{13}$CO(1-0) molecular transitions, we detected four components with kinematic velocities \mbox{$v_{\text{lsr}}=48$\,km/s} (\textit{R1a}), \mbox{$v_{\text{lsr}}=41$\,km/s} (\textit{R1c}), $v_{\text{lsr}}=58$\,km/s (\textit{R1d}), 
and \mbox{$v_{\text{lsr}}=67$\,km/s} (\textit{R1e}).
CS(1--0) was also found  in \textit{R1a,b,c}. However, only the component \textit{R1a} and \textit{R1b} remained  visible in NH$_{3}$(1,1).
The CS(1--0) emission in \textit{R1a} ($v_{\text{lsr}}$=48 km/s, see Table \ref{fitlabel}) is quite broad ($\Delta v_{\text{FWHM}}\sim 10$ km/s) compared to the other fainter components along the line of sight.

We also found two  44GHz CH$_3$OH masers that we labeled \textit{CH1} \mbox{($v_{\text{lsr}}=$46 km/s)} and \textit{CH2} ($v_{\text{lsr}}=$56 km/s) (see Fig. \ref{CH3spitzer}). They are thus associated to the molecular cloud traced by the component
\textit{R1a} and \textit{R1d}.
While the component \textit{CH1} seems connected to the IR bubble \textit{N22} (labeled by \citealt{Churchwell2006}), \textit{CH2} is likely associated with the H\textsc{ii}  region HDRS G018.097-0.324 (\textit{H2} in Fig.\,\ref{CH3spitzer}).
We also found H51$\alpha$, H62$\alpha$, H65$\alpha$, and H69$\alpha$ emission at $v_{\text{lsr}}\sim 50-55$ km/s coincident with region \textit{H1} at kinematic distances roughly agreeing  with the distance $d$=4.3\,kpc proposed by \citet{JacksonGRS}.
From the H\textsc{I} 21 cm SGPS \citep{SGPS2005}, we observed several absorptions features towards the region \textit{H1} coincident with the CO(1--0) emission in \textit{R1a} while no H\textsc{i} absorption was associated with \textit{R1e}. (see Fig. \ref{HIIspectra}).
Consequently, the cloud is positioned in the near distance $d$=3.9 kpc.

Fig. \ref{CH3spitzer} also shows the CS(1--0) and NH$_3$ integrated emission between \mbox{$v_{\text{lsr}}=40-60$\,km/s} and the recombination line H62$\alpha$ between \mbox{$v_{\text{lsr}}=45-65$\,km/s}.

\begin{figure*}
  \begin{minipage}{\textwidth}
   \includegraphics[width=\textwidth,angle=0]{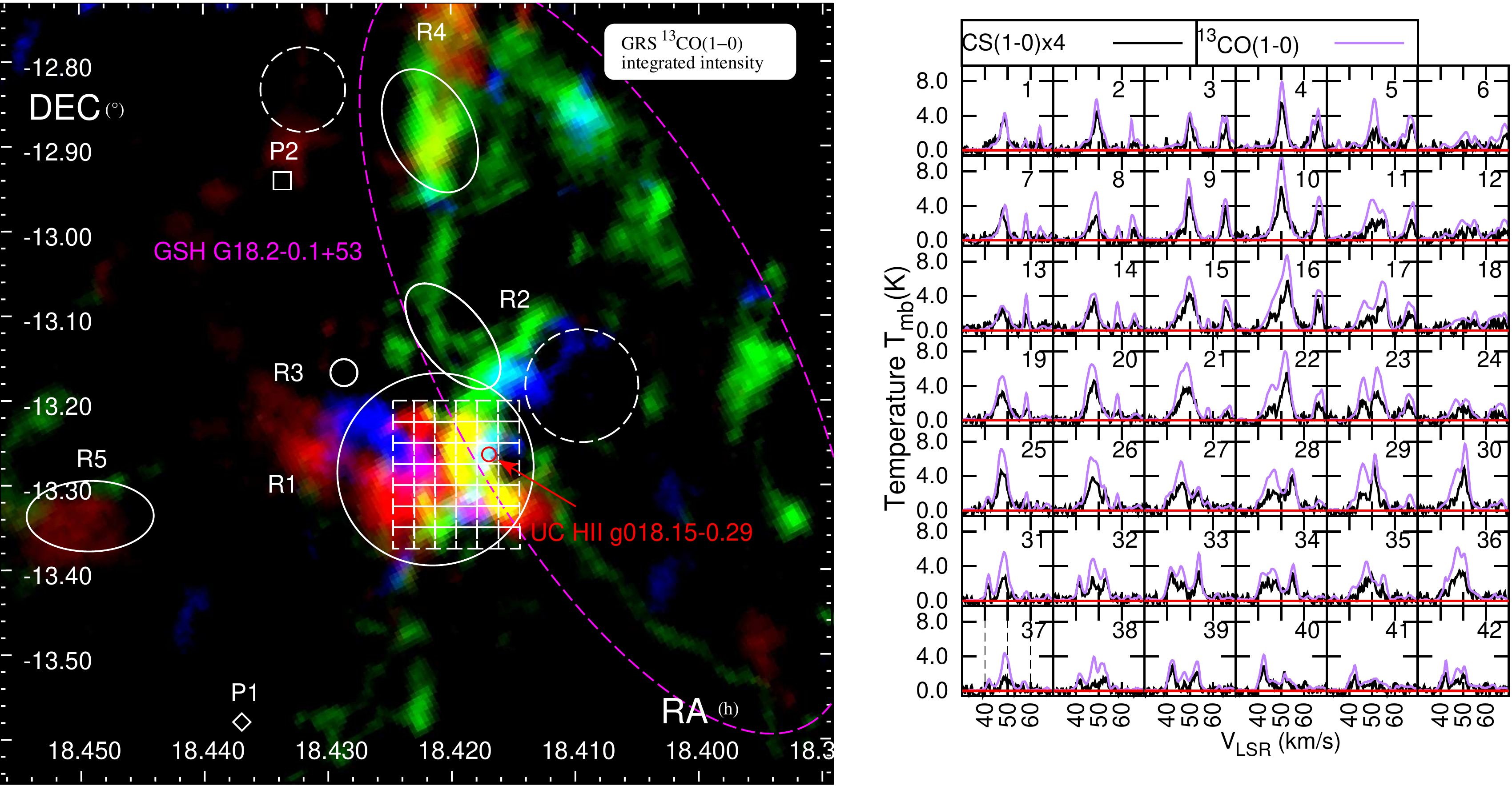}%
  \caption{\textit{(Left)} Three-colour image illustrating  the GRS $^{13}$CO(1--0) towards HESS\,J1826$-$130 integrated intensity at three velocity ranges : $45-50$\,km/s (red), $50-55$\,km/s (green), $55-60$\,km/s (blue). The H\textsc{ii}  region UC H\textsc{ii}  G018.15-0.29
  is  indicated in red. The purple dashed ellipse indicates the position and size of the putative molecular shell GSH\,18.1-0.2+53 \citep{Paron2013}.
  \textit{(Right)} Spectra of GRS $^{13}$CO(1--0) (purple lines in colour version) and Mopra CS(1--0) emission (black lines) averaged over the boxes shown in the left panel. } 
  \label{GRSinteg}
 \end{minipage}
\end{figure*}

We noted that the NH$_3$(1,1) appeared less prominent away from the region \textit{H1}, \textit{N21} and \textit{N22} as opposed to the CS(1--0).
Although the CS(1--0) emission appears uniform across \textit{R1} between $v_{\text{lsr}}=40-60$\,km/s, the spectral lines averaged over the grid of boxes as shown in Fig. \ref{GRSinteg} reveals several contiguous cloud sub-components.
For instance, the CS(1--0) emission indicated many line shape variations towards the south of \textit{R1}, with several peaks with small velocity separations (e.g boxes 32 to 35), and rapid variations
of the peak velocity (e.g boxes 13 to 17).
Additionally, Fig.\,\ref{NR1vel} also indicates that the NH$_3$(1,1) and CS(1$-$0) emission located at $v_{\text{lsr}}=45-60$ km/s broad velocity structures to the CS and NH$_3$ compared to the emission in \textit{R1e}.

The $^{13}$CO(1--0) spectral lines illustrated similar features.
Interestingly, from the three colour map in Fig.\,\ref{GRSinteg} showing the GRS $^{13}$CO(1--0) integrated intensity between \mbox{$v_{\text{lsr}}=45-50$\,km/s} (red), \mbox{$v_{\text{lsr}}=50-55$\,km/s} (green), 
\mbox{$v_{\text{lsr}}=55-60$\,km/s} (blue), we observed that the structure shown in red was distinct from the arc-shaped structure displayed in green.
We also noted a spatial overlap between all emission across the  aforementioned velocity bands next to the H\textsc{ii} region G018.15-0.29 which suggests  there is a physical link between the H\textsc{ii}  region and the 
these structures.
The presence of the double-peaked emission found in boxes 17 and 23 ($v_{\text{lsr}}$=46 km/s and $v_{\text{lsr}}$=56 km/s) which differs from the single-peaked emission found in box 5 \mbox{($v_{\text{lsr}}$=51 km/s)} may be caused by the presence 
of this continuum source. Consequently, it is likely that the component \textit{R1a} and \textit{R1d} are physically connected.

The physical parameters listed in Table \ref{MASStot}.a in the appendix shows that the molecular gas traced by the \textit{R1a} component is much more massive than the gas traced by the other components in the line of sight.
From our CS and CO analysis, we found \mbox{$M_{\text{H}_2}$(CS)=$1.0\times 10^5\solmass$} and \mbox{$M_{\text{H}_2}\left(\text{CO}\right)=1.2\times 10^5\solmass$} respectively which is within the Virial mass range 
\mbox{$M_{\text{vir}}=0.5-2.1\times10^{5}\solmass$},
and averaged densities \mbox{$n_{\text{H}_2}$(CS)=$7.5\times10^{2}$ cm$^{-3}$} and \mbox{$n_{\text{H}_2}\left(\text{CO}\right)=9.6\times 10^{2}$ cm$^{-3}$}.
The similar mass estimation from our CS and CO analysis may suggest the observed molecular gas are concentrated in clumps of density roughly equal to the CS(1--0) critical density \mbox{$n_{\text{c}}$\,=\,$2\times 10^{4}$ cm$^{-3}$}.
However, a lower mass and density estimates were attained with our NH$_3$ analysis with  \mbox{$M_{\text{H}_2}=1.4\times10^4\solmass$} and \mbox{$n_{\text{H}_2}=2.0\times10^{2}$ cm$^{-3}$}.

\subsubsection{Region R2}

Towards region \textit{R2} (RA=18.420h, Dec=$-13.125^{\circ}$), we detected three CO(1--0) components with velocity  \mbox{$v_{\text{lsr}}=$51 km/s} (\textit{R2a}), \mbox{$v_{\text{lsr}}=$43 km/s} (\textit{R2c}), $v_{\text{lsr}}=68$ km/s (\textit{R2e}) (see Table \ref{fitlabel}). 
However, $^{13}$CO(1--0), CS(1--0) and NH$_3$(1,1) were solely found in \textit{R2a} and \textit{R2e}.
We also observed SiO$_{\text{v}=0}$(1--0) emission, whose centroid velocity coincided with the component \textit{R2e} revealing the presence of post-shocked gas (see Fig. \ref{CH3spitzer}). The combined detection of a 44GHz maser
CH$_3$OH(7$_1$--6$_0$) (region \textit{CH3}), NH$_3$(3,3) (see Fig.\,\ref{NH2233}) and cyanopolyyne HC$_3$N (region \textit{HC3}, see Fig.\,\ref{spitzerhCCN}) and their spatial connection with the IR 
source (see Fig. \ref{CH3spitzer} left panel) suggested the shock may have been caused by outflows from nearby star forming regions.

Our CS and NH$_3$ analysis indicated that this molecular cloud was optically thick ($\tau_{\text{CS(1--0)}}$=1.5 and $\tau_{\text{NH}_3\text{(1,1)}}$=3.5). Consequently, the CO(1--0) emission may suffer strong optical depth
effects which would cause line width broadening ($\Delta v$=12 km/s, see Table \ref{fitlabel}).

Assuming the molecular gas is located at d=4.6 kpc (near distance), we obtained the following masses $M_{\text{H}_2}\left(\text{CS}\right)=3.4\times10^{4}\solmass$, $M_{\text{H}_2}\left(\text{NH}_3\right)=8.1\times10^{4}\solmass$ and \mbox{$M_{\text{H}_2}\left(\text{CO}\right)=2.9\times10^{4}\solmass$}
which were all within the Virial mass limits $M_{\text{vir}}=0.3-1.1\times10^{5} \solmass$.
We also determined the averaged densities \mbox{$n_{\text{H}_2}\left(\text{NH}_3\right)\sim 4.8\times 10^3$ cm$^{-3}$},$n_{\text{H}_2}\left(\text{CS}\right)\sim 2.0\times 10^3$ cm$^{-3}$ and $n_{\text{H}_2}\left(\text{CO}\right)\sim 1.7\times 10^3$ cm$^{-3}$.

The component \textit{R2a} is associated  to the component \textit{R1a} (see Fig. \ref{CSCOGRS}) and the masses obtained are : $M_{\text{H}_2}\left(\text{CS}\right)>3.4\times10^{3} \solmass$, $M_{\text{H}_2}\left(\text{NH}_3\right)>4.5\times10^{2}\solmass$
 and $M_{\text{H}_2}\left(\text{CO}\right)=1.3\times10^{4}\solmass$, and densities $n_{\text{H}_2}\left(\text{CS}\right)>3.3\times10^{2}$ cm$^{-3}$, $n_{\text{H}_2}\left(\text{NH}_3\right)>4.3\times10^{1}$ cm$^{-3}$
 and \mbox{$n_{\text{H}_2}\left(\text{CO}\right)=\,1.2\times10^{3} $cm$^{-3}$} (see Table \ref{MASStot}.b).
The small fraction of dense gas detected by the CS and NH$_3$ tracers at \mbox{$v_{\text{lsr}}\sim 50$\,km/s} explains the large discrepancies between the different masses and densities.

Finally, our CO analysis revealed the H$_2$ mass traced by the component \textit{R2c} is $M_{\text{H}_2}=8.8\times10^{3}\solmass$ and a density $n_{\text{H}_2}\left(\text{CO}\right)=3.9\times10^{2}$ cm$^{-3}$.

\subsubsection{Region R3}
We detected three spectral components inside \textit{R3} (RA,Dec)=(18.429h, $-13.178^{\circ}$).
The two components \textit{R3a} and \textit{R3e} from CO(1--0) and $^{13}$CO(1--0) observations appear to be extensions of the molecular gas found in the regions \textit{R1} and \textit{R2}, and
 their masses listed in Table \ref{MASStot}.c indicate small mass contribution.
On the other hand, the component \textit{R3f}  \mbox{($v_{\text{lsr}}\,=\,43$ km/s)} showed  prominent NH$_3$(1,1) and CS(1$-$0) emission.
The additional detection of the cyanopolyyne HC$_3$N(5--4,F=4--3) suggested the molecular cloud may be associated with the H\textsc{ii}  regions G018.303-0.389 and G018.305-0.392 \citep{White2005}.
From the H\textsc{i} spectral lines shown in Fig. \ref{HIIspectra}, all CO(1--0) emission found in \textit{R3} was associated with an H\textsc{I} absorption feature and consequently puts the molecular cloud in the 
far distance of $d$=13.4\,kpc.
Assuming the gas traced by the component \textit{R3f} has an angular size equal to the Mopra NH$_3$(1,1) beam size, we obtained the following H$_2$ masses $M_{\text{H}_2}\left(\text{CS}\right)=1.5\times10^{5}\solmass$, 
$M_{\text{H}_2}\left(\text{NH}_3\right)=1.3\times10^{4}\solmass$, $M_{\text{H}_2}\left(\text{CO}\right)=1.5\times10^{4}\solmass$ and Virial mass $M_{\text{vir}}=0.7-2.5\times10^{4}\solmass$ which gives the following H$_2$ densities 
$n_{\text{H}_2}\left(\text{CS}\right)=7.6\times10^{3}$ cm$^{-3}$, $n_{\text{H}_2}\left(\text{NH}_3\right)=6.2\times10^{2}$\,cm$^{-3}$, $n_{\text{H}_2}\left(\text{CO}\right)=9.3\times10^{2}$\,cm$^{-3}$.

\subsubsection{Region R4}
Region \textit{R4}  located north of HESS\,J1826$-$130, is nearby the supernova remnant SNR\,G018.6$-$0.2.
From NH$_3$(1,1) and CS(1--0) observations, we detected two components \textit{R4a} and \text{R4b} with small velocity separation ($v_{\text{cent}}$=45 km/s and $v_{\text{cent}}$=51 km/s).

As shown in Fig.\,\ref{NR1vel}, we observed no CS and NH$_3$ emission connecting the components \textit{R4a} and \textit{R4b}.
From Fig.\,\ref{GRSinteg}, we noted that the molecular gas in \textit{R4a} (shown in green) appeared to be part of the putative molecular shell GSH\,G018.2--0.1+53 suggested by \citet{Paron2013}
 whereas the gas in \textit{R4b} (in red) appeared isolated.
 However, their similar morphologies as shown by the overlap of the two components (in yellow) the similarities of the $^{13}$CO(1--0), CS, and NH$_3$ spectral lines may indicate some association.

Assuming \textit{R4a} and \textit{R4b} were situated at the near distance, 
we derived the total mass lower limit for \textit{R4a} \mbox{$M_{\text{H}_2}\left(\text{CS}\right)>7.4\times10^{3}\solmass$}, $M_{\text{H}_2}\left(\text{NH}_3\right)>2.0\times10^{3} \solmass$, $M_{\text{H}_2}\left(\text{CO}\right)=1.6\times10^{4}\solmass$  and 
\mbox{$M_{\text{H}_2}\left(\text{CS}\right)>5.0\times10^{3}\solmass$}, $M_{\text{H}_2}\left(\text{NH}_3\right)>1.2\times10^{3}\solmass$ and \mbox{$M_{\text{H}_2}\left(\text{CO}\right)=1.1\times10^{4}\solmass$} for \textit{R4b} agreeing with their Virial mass ranges $M_{\text{vir}}=0.7-2.5\times10^{4}\solmass$ and 
\mbox{$M_{\text{vir}}=0.6-2.2\times10^{4}\solmass$} respectively (see Table\,\ref{MASStot}.d).
From CO analysis, we obtained the densities \mbox{$n_{\text{H}_2}\left(\text{CO}\right)=9.3\times10^{2}$cm$^{-3}$} and \mbox{$n_{\text{H}_2}\left(\text{CO}\right)=8.2\times10^{2}$\,cm$^{-3}$} for \textit{R4a} and \textit{R4b} respectively. 
In the case where \textit{R4a} and \textit{R4b} where associated to the same molecular complex at d$\sim$4\,kpc , the total mass obtained would be $M_{\text{H}_2}\left(\text{CO}\right)\sim3.0\times10^{4}\solmass$. 

Therefore, although the region \textit{R4} is unlikely to be physically related to HESS\,J1825$-$137 and HESS\,J1826$-$130, it highlights the complexity of the structure of the molecular gas in the line of sight.

\subsubsection{Region R5}
The region \textit{R5} is located 20$^{\prime}$ away from the pulsar PSR\,J1826-1334.
From CO(1--0) observations, three components in region \textit{R5} with centroid velocity $v_{\text{cent}}$=51 km/s (\textit{R5a}), $v_{\text{cent}}$=45 km/s (\textit{R5b}), $v_{\text{cent}}$=61 km/s (\textit{R5d}) were detected.
However, $^{13}$CO(1--0), CS(1--0) were solely observed in \textit{R5a} and \textit{R5b} and weak NH$_3$(1,1) emission ($<3T_{\text{rms}}$) was found only in \textit{R5a}. 

From Fig. \ref{NR1vel}, it appears that most of the emission is located between $v_{\text{lsr}}=45$\,km/s and $v_{\text{lsr}}=48$\,km/s.

We derived H$_2$ masses of \mbox{$M_{\text{H}_2}\left(\text{NH}_3\right)>2.0\times10^{3}\solmass$}, $M_{\text{H}_2}\left(\text{CS}\right)>7.4\times10^{2}\solmass$, $M_{\text{H}_2}\left(\text{CO}\right)=9.5\times10^{3}\solmass$ for \textit{R5a}
 and $M_{\text{H}_2}\left(\text{CS}\right)>1.8\times10^{3}\solmass$ and $M_{\text{H}_2}\left(\text{CO}\right)=1.1\times10^{4}\solmass$ for \textit{R5b}. Therefore, the  molecular gas traced by \textit{R5a} and \textit{R5b} appears less clumpy as opposed to \textit{R1a}.
  From CO(1--0) analysis, we obtained the densities \mbox{$n_{\text{H}_2}\left(\text{CO}\right)=6.8\times10^{2}$ cm$^{-3}$} and $n_{\text{H}_2}\left(\text{CO}\right)=6.1\times10^{2}$ cm$^{-3}$ for \textit{R5a} and \textit{R5b} respectively (see Table\,\ref{MASStot}.e).
  Therefore, the molecular clouds located inside \textit{R5} are marginally denser than our other studied regions.
   If \textit{R5a} and \textit{R5b} were to be physically connected and located at \mbox{$d$=4 kpc}, we would obtain the following total mass \mbox{$M_{\text{H}_2}\left(\text{CO}\right)=2.3\times10^{4}\solmass$}.\

\subsubsection{Region R6}
The region \textit{R6} is located in the southern part of HESS\,J1825-137. It is surrounded by several H\textsc{ii}  regions  and the SNRs G017.4$-$0.1 and G017.0$-$0.0
We found CO and $^{13}$CO(1-0) emission with a centroid velocity at \mbox{v$_{\text{lsr}}\sim$ 44 km/s}.
As shown in Fig. \ref{SPEC}, we noted that the CO(1--0) emission and its isotopologue $^{13}$CO(1--0) revealed a broad positive wing \mbox{($v_{\text{lsr}}=45-55$\,km/s)}.
However our deep pointing in CS(1--0) revealed no such features.

From our deep pointing measurements, we obtained \mbox{$N_{\text{H}_2}=7.5\times 10^{21}$ cm$^{-2}$}. If we assumed the molecular clouds have uniform column density across the molecular clouds, we would obtain 
a total mass \mbox{$M_{\text{H}_2}\left(\text{CS}\right)>3.6\times10^{4} \solmass$}(see Table\,\ref{MASStot}.f), which would be a factor of two smaller than the mass derived using our Nanten CO(1--0) observations \mbox{$M_{\text{H}_2}$(CO)=$7.6\times10^{4}\solmass$}.

Region \textit{R6} and the surrounding molecular gas reveals a broad spatial distribution of $^{13}$CO(1--0) and the molecular cloud may be associated to the TeV source HESS\,J1825$-$137.
This molecular cloud  may therefore influence the morphology of the south region of the TeV PWN.

\subsection{Summary}
We detected several molecular clouds along the line of sight.
Notably, we observed a small molecular cloud located at \mbox{$v_{\text{lsr}}$=18\,km/s} overlapping with the pulsar PSR\,J1826-1256 and whose kinematic distance $d=1.7$\,kpc roughly agreed with the pulsar's predicted 
distance $d$=1.4 kpc from \citet{Wang}.

The molecular gas located at \mbox{$v_{\text{lsr}}=46-55$\,km/s} and matching the pulsar's distance have a mass \mbox{$M_{\text{H}_2}$=$3.3\times 10^5 \solmass$} where $\sim 30$\% resides inside the region \textit{R1}.
Moreover, we observed prominent and extended CS(1--0) and NH$_3$(1,1) emission in \textit{R1a} and thus it suggests that the observed molecular gas consists of dense clumps likely to exceed the CS(1--0)
critical density \mbox{$n_c\sim2\times10^4$ cm$^{-3}$} at 10 K.

The molecular gas traced by the component \textit{R1a} also revealed complex CS and $^{13}$CO(1--0) spectral lines surrounding the H\textsc{ii}  region G018.15-00.29 and towards HESS\,J1825$-$137 with several intensity peaks
with little velocity separations and rapid variations of the velocity peaks.

 We also remarked that the components found at \mbox{$v_{\text{lsr}}=44-46$\,km/s} `b' often overlapped and shared similar properties with the component `a' at $v_{\text{lsr}}=46-55$\,km/s which may indicate possible physical
connection.

Finally the H\textsc{I} and $^{13}$CO(1--0) data showed the presence of a plausible void centred at the SNR\,G018.6$-$0.2 at \mbox{$v_{\text{lsr}}\sim60-65$\,km/s} and associated with the molecular cloud where the dense gas traced by the component \textit{R2e}
resided. This suggests that this SNR may be located at a distance $d$=4.6\,kpc (near distance) or $d$=11.4\,kpc (far distance).

\section{Discussion}
\label{section5}
\subsection{Dynamics of the dense molecular gas in region \textit{R1a}-Looking for the progenitor SNR}
\begin{figure}
  \includegraphics[width=0.5\textwidth]{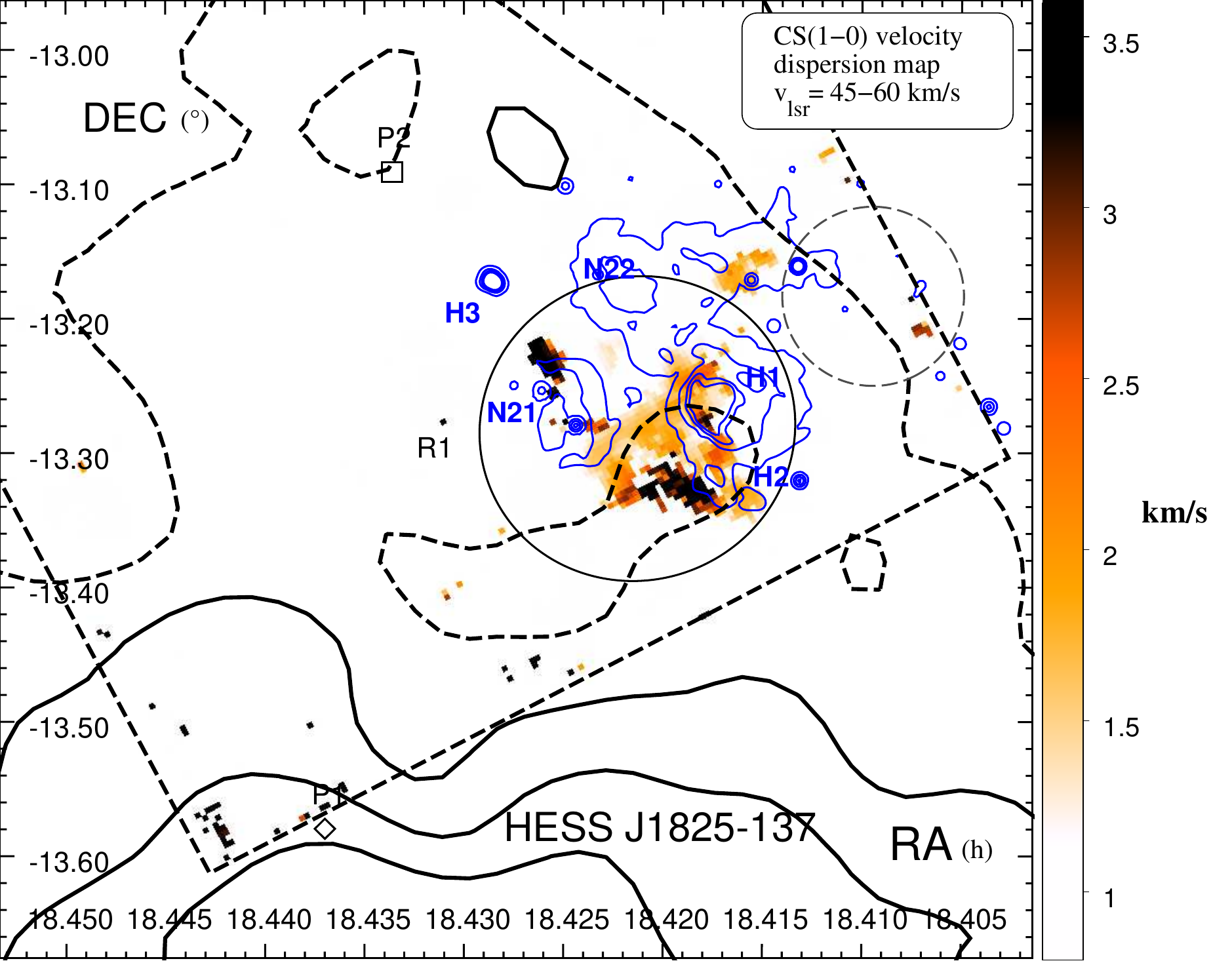}
\caption{CS(1--0) velocity dispersion $v_{\text{disp}}$ map between $v_{\text{lsr}}=45-60$\,km/s with $T_{A}^{*}>4 T_{\text{rms}}$, overlaid
by the 8$\mu m$ Spitzer GLIMPSE contours (blue in colour version).} 
\label{csvelocity}
\end{figure}
We focus now  on the gas dynamics to probe the level of disruption in the observation of this region structure of \mbox{CS(1--0)}. Bubbles, core-collapsing clouds and shocks are the common causes of gas disruption.
We have shown that the dense molecular cloud traced by the component \textit{R1a} displays complex morphology and spectral line profiles.
The velocity dispersion, or second moment, map of CS(1--0) in  Fig. \ref{csvelocity} indicates broad dense gas overlapping with the H\textsc{ii}  region UC H\textsc{ii}\,G018.15-0.28.

Most of the CS(1--0) emission towards the centre of \textit{R1a}  displays a mild velocity dispersion ($v_{\text{disp}}\sim 2$ km/s). There is also no appreciably 
 broad gas overlapping the IR bubbles \textit{N21} and \textit{N22}. 
However, we observe a $\sim 3.5-4$\,km/s velocity dispersion to the south of \textit{R1a} towards HESS\,J1825$-$137 which does not seem related to any IR emission as shown in Figs. \ref{csvelocity}.

\begin{figure}
 \includegraphics[width=0.5\textwidth]{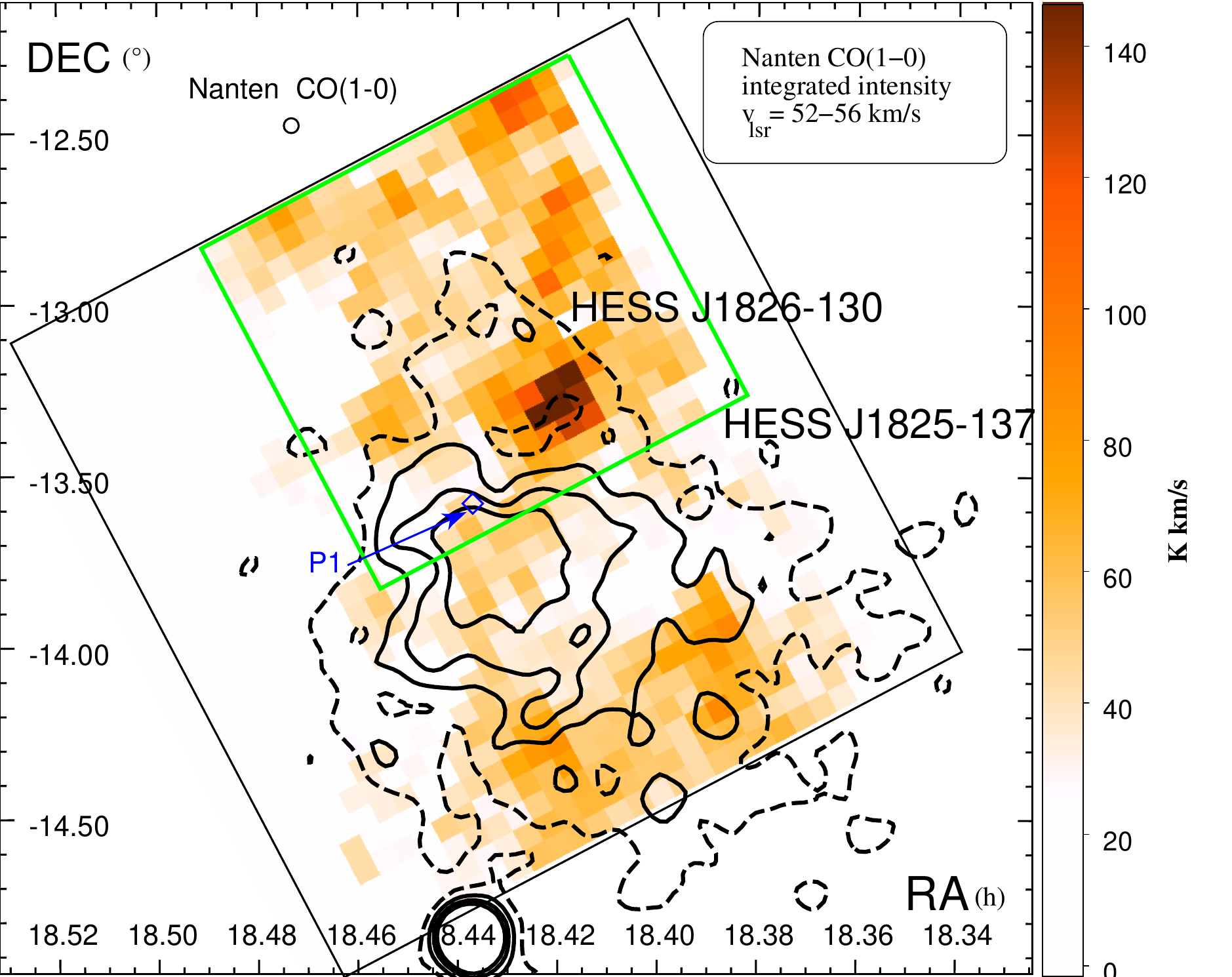}
 \caption{Nanten CO(1--0) integrated intensity between $v_{\text{lsr}}=52-56$ km/s overlaid by HESS TeV contours. The large black square represents the region where the overall averaged ambient density $n_\text{amb}$ was estimated (see section 4.1) with the CO(1--0) emission within 
the green box (in colour version) being excluded.}
\label{cooling1}
\end{figure}

We now focus on potential causes of the broad CS(1--0) lines in \textit{R1a}.
Little is known about the progenitor SNR of HESS\,J1825$-$137.
From hydrodynamical simulation of the PWN, the radius of the SNR is expected to be four times the radius of this middle-aged PWN $r_{\text{SNR}}\sim 4r_{\text{PWN}}$ \citep{swaluw}. 
With the TeV radius of the PWN being $r_{\text{PWN}}\sim35$ pc \citep{Aha2006}, the predicted radius of the SNR thus becomes significantly large at $r_{\text{SNR}}\sim140$ pc.

\citet{DeJager} indicated that the SNR's radius could reach \mbox{$r_{\text{SNR}}\sim 120$ pc} if they assumed a kinetic energy \mbox{$E_{\text{SN}}=3\times10^{51}$ erg} from the supernova shock, an ambient density
\mbox{$n_{\text{amb}}=0.001$ cm$^{-3}$}, and an age of the system  \mbox{$t_{\text{age}}\sim 40$ kyr}.
By taking the Nanten CO(1--0) emission over a narrow $v_{\text{lsr}}$ range (52$-$56 km/s) centred on the pulsar PSR\,J1826--1334 distance and discarding the contribution from the dense region inside the green box in Fig.\,\ref{cooling1},
 we obtain a density \mbox{$n_{\text{H}}\sim 4$ cm$^{-3}$.} 
However, the typical noise level $T_{\text{rms}}$/ch$\sim0.4$\,K of the Nanten CO(1--0) data \citep{Fukui} greatly affects our density estimation which should then only be used as a upper-limit.
Thus, our derived density does not at the moment refute the ambient density predicted by \citet{DeJager}.
Besides this, due to the contamination of the CO(1--0) emission resulting from local and distant kinematic components, it also becomes difficult to identify the presence of the void caused by the SNR's progenitor star. The better sensitivity and velocity resolution of the NASCO \citep{NASCO1} or Nobeyama\footnote{http://www.nro.nao.ac.jp/\~nro45mrt/html/index-e.html} surveys 
 of the CO and $^{13}$CO emission may provide solutions to this issue.

\begin{figure}
 \includegraphics[width=0.5\textwidth]{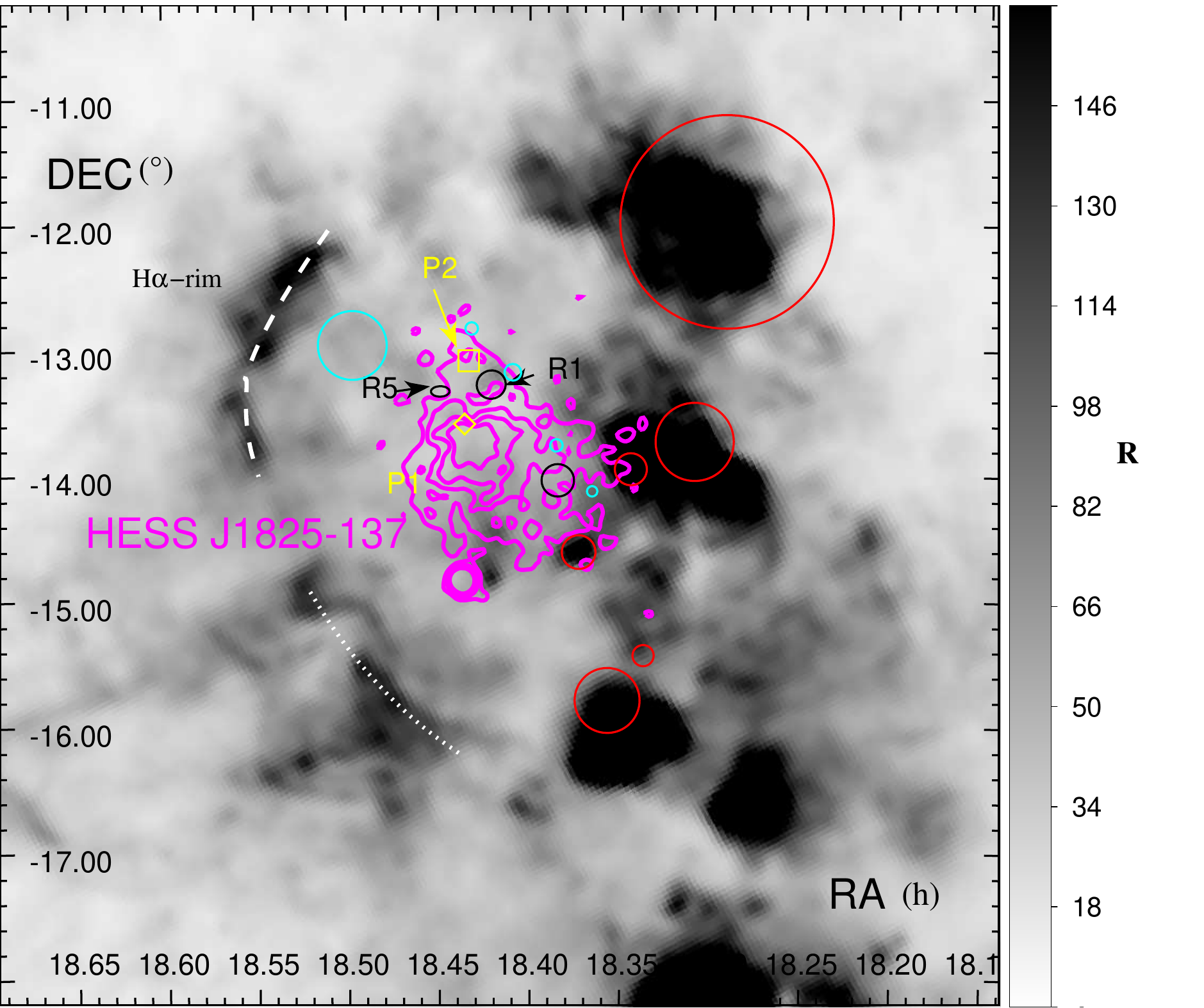}
 \caption{H$\alpha$ image towards HESS\,J1825-137 in Rayleigh units overlaid by the HESS contours in magenta.
 The two pulsars labeled P1 and P2 are shown in yellow while the SNRs are displayed in cyan circles (see online version for colours).
 The white dashed lines represent the SNR H$\alpha$ rim reported by \citet{Stupar2008} while the white dotted lines indicate another putative H$\alpha$ rim which might be associated with the 
 HESS\,J1825$-$137's progenitor SNR. The red circles encompassing the strong H$\alpha$ emission on the right hand side are catalogued by \citet{Anderson2014} as H\textsc{ii}  regions.}
 \label{Halpha}
\end{figure}
Interestingly though, \citet{Stupar2008} reported a H$\alpha$ rim (see white dashed lines in Fig.\,\ref{Halpha}) located roughly 120 pc away from PSR\,J1826-1334 (assuming the H$\alpha$ rim is located at $d=4$ kpc) and with  a ratio S\textsc{ii}/H$\alpha\sim1.33$ typical 
of SNR shock.
Moreover, based on the sharp gradient in H$\alpha$, we might also speculate that the reported H$\alpha$ rim is also seen  to the south-east of HESS\,J1825$-$137 as shown 
in white dotted line in Fig.\,\ref{Halpha}.
The strong H$\alpha$ emission to the west of HESS\,J1825$-$137 (red circles) has been catalogued as H\textsc{ii} regions \citep{Anderson2014}.
The projected separation of the H$\alpha$ rim roughly matches with the SNR's expected radius based on $R_{\text{SNR}}/R_{\text{PWN}}\sim4$ suggested by \citet{DeJager}. 
Furthermore, the lack of H$\alpha$ emission north of HESS\,J1825$-$137 may arise due to the blocking effect of dense cloud responsible for the crushed PWN. 

The possible association between the H$\alpha$ rim and the TeV source could provide important information about the environment surrounding HESS\,J1825$-$137.

We now consider whether this SNR would contribute to the turbulence found inside \textit{R1a}.
We do not observe direct evidence of post-shocked gas inside at $v_{\text{lsr}}=45-60$\,km/s such as SiO(1--0,v=0) emission or catalogued OH\,1720 MHz masers. 
Nonetheless, if the shock did reach the cloud, we do expect the shock speed $v_{\text{s}}$ to be considerably lowered due to the high averaged density found in this region.
By applying Eq.\,10 from \citet{Uchiyama2010} :
 \begin{equation}
  v_{\text{s}}\approx65 \left(\frac{n_{\text{R1a}}}{100\text{ cm}^{-3}}\right)^{-1/2}\left(\frac{E_{\text{SN}}}{10^{51}\text{ erg}}\right)^{1/2}\left(\frac{R_{\text{R1a}}}{12.5\text{ pc}}\right)^{-3/2} \text{ km/s}
  \label{Uchiyamaeq}
  \end{equation}
 
 and assuming the distance to \textit{R1a} boundary  \mbox{$R_{\text{R1a}}=20$\,pc}, a proton density found from our CO analysis \mbox{$n_{\text{R1a}}=2.7\times10^{3}$\,cm$^{-3}$},
 we find that the shock could reduce to a speed \mbox{$v_{\text{s}}\sim 10$\,km/s}. Consequently, using the age upper-limit  \mbox{$t=40$\,kyr} for the SNR, the shock would have only travelled less than 1\,pc (i.e an angular distance $\theta\sim0.01^{\circ}$)  inside the dense molecular cloud.
 Therefore, the SNR might only contribute to the disruptions found south of \textit{R1a} (see Fig.\,\ref{csvelocity}) where we see a broader velocity dispersion $v_{\text{disp}}$.
 
Alternatively, the broad velocity dispersion may also be caused by two distinct but contiguous velocity components as shown in Fig.\,\ref{GRSinteg} (red and green components) and thus may not be an indicator of randomly distributed disruption in this region.
In fact, we note that this cloud shows similarities with the studied molecular clouds next to the Serpens cluster \citep{Duartecabral2011} and RCW120 \citep{Torii2015}, 
whose velocity components are thought to be caused by cloud-cloud collisions.
Cloud-cloud collisions, studied using hydrodynamical simulations (\citealt{Cloudcollisiontheory}, \citealt{Duartecabral2011},\citealt{Takahira2014},\citealt{Torii2015}), recently renewed popularity  to explain the presence
 of high mass star-formation inside molecular clouds.
 Notably, such collisions between  molecular clouds are thought to generate OB stars, filamentary clouds, dense cores and complex velocity distribution. 
 In our region of study, \citet{Paron2013} in fact detected several O and B stars next to the IR bubble N22, N21 and the UC H\textsc{ii}\,G018.15-00.28 towards
 the molecular gas \textit{R1a} component.

 \subsection{TeV emission of HESS\,J1826--130}
 An important question is whether or not HESS\,J1826$-$130 can be associated with HESS\,J1825$-$137.  
 Now, we will discuss potential origins of this northern TeV emission.
 For HESS\,J1826$-$130, \citet{Deil2015} reported its location at (18.434,$-13.02^{\circ}$), and a TeV flux above 1 TeV $F\left(>1\text{ TeV}\right)=7.4\times10^{-13}$\,ph\,cm$^{-2}$\,s$^{-1}$.

 \subsubsection{CRs from the progenitor SNR of PSR\,J1826--1334}
 
The general spatial match between the molecular cloud and the TeV emission of HESS\,J1826$-$130 may suggest a hadronic origin. Here, the progenitor SNR of HESS\,J1825$-$137 is an obvious candidate source for CRs in the region.
A key question is whether the observed emission can be explained by the {\em sea} (Galactic plane averaged CR energy density \mbox{$w_{\text{CR}}\sim 1$\,eV\,cm$^{-3}$})  of cosmic-rays or require the presence of a nearby CR source.
Using Eq.\,10 from \citet{Aha1991} and the mass of the molecular cloud \mbox{$M_{\text{H}_2}=3.3\times10^{5}\solmass$} calculated earlier, 
 we obtain \mbox{$F\left(>1\text{ TeV}\right)=5.8\times10^{-14}$\,ph\,cm$^{-2}$\,s$^{-1}$} produced by the sea of cosmic-rays towards HESS\,J1826$-$130.
 This predicted flux is $\sim$15 times below the observed flux estimated above.
 Therefore a nearby  accelerator providing CRs at an energy density 15 times the Earth-like value is required.
 The required CR density enhancement to produce the observed TeV flux towards HESS\,J1826$-$130 can easily be reached by SNRs (e.g W28; see \citealt{Aha2008W28}).
 Therefore, the progenitor SNR of PSR\,J1826$-$1334 may contribute to the HESS\,J1826$-$130 TeV emission.

 \subsubsection{Other potential particle accelerators}
 \citet{Paron2013} argued the distance of SNR\,G018.2$-$0.1 to be $d=4$\,kpc based on its possible association with the nearby H\textsc{ii} regions. Its small projected radius $r\sim$4\,pc
 would imply a very young SNR with age $<1000$\,yr.
 By comparing our CO(1--0) column density the those derived from X-ray measurements \mbox{$N_{\text{H}}=7.2\times10^{22}$\,cm$^{-2}$} \citep{Sugizaki2001} and \mbox{$N_{\text{H}}=5.7\times10^{22}$\,cm$^{-2}$} \citep{Leahy2014}, 
 we would argue an SNR distance greater than 4 kpc, consistent with the distance estimate from \citet{Leahy2014}.
 Additionally, \citet{sigmaDratio2014} suggest a much further distance $d>8.8$\,kpc based on their updated $\Sigma-D$ relation.
 In the case where the SNR is located at $d=4$\,kpc, CRs would probably remain confined inside the SNR shock and would not be responsible of the TeV $\gamma$-ray emission found in HESS\,J1826$-$130.
 We did not find molecular gas overlapping the HESS\,J1826$-$130 emission in the case where $d>4$\,kpc.
 Therefore, we suggest that the TeV $\gamma$-ray emission towards HESS\,J1826$-$130 cannot be produced by CRs accelerated by the SNR\,G018.2$-$0.1.
 
 The radio-quiet pulsar PSR\,J1826$-$1256 (P2), which powers the diffuse X-ray nebula PWN\,G018.5$-$0.4 \citep{Mallory2001GeV,Robert}, may also be a candidate for the origin of HESS\,J1826$-$130.
 The distance $d=1.2-1.4$\,kpc suggested by \citet{Wang}  infers a TeV $\gamma$-ray efficiency
 $\eta_{\gamma}\sim2\times10^{-4}$ which is at the lower end of typical $\eta_{\gamma}$ values for $\dot{E}_{\text{SD}}\sim10^{36-37}$\,erg\,s$^{-1}$ \citep{kargaltsev2}.
 We find that the Nanten CO(1--0) emission located at this distance ($v_{\text{lsr}}=10-25$\,km/s, see Fig.\,\ref{NantenCO1025}) overlaps  the pulsar and makes the PWN scenario at this distance unlikely.
 However, the adjacent position of the molecular gas at \mbox{$v_{\text{lsr}}=60-80$\,km/s} (with near/far distance $d=4.6/11.4$\,kpc) appears spatially consistent with the PWN scenario.
 The molecular cloud north of P2 indeed support the offset position of the pulsar (coincident with AX\,J1826.1$-$1257, see \citealt{Ray1}) with respect to the X-ray \citep{Robert} and TeV emission.
 At these distances, $\eta_{\gamma}$ would then rise to $1.5\times10^{-3}$ ($d=4.6$\,kpc) and $1.0\times10^{-2}$ ($d=11.4$\,kpc), more consistent with the canonical $\eta_{\gamma}$ values.
 
 The plausible shell at \mbox{$v_{\text{lsr}}=60-65$\,km/s} spatially coincident with the SNR\,G018.6$-$0.4 puts this SNR  at the same distance to the pulsar P2 
  and may suggest an association between the two objects.
  To reconcile the small size of SNR\,G018.6$-$0.2 and the characteristic age of P2 ($\tau_c=13$\,kyr), they need to be placed at a far distance $\sim 11.4$\,kpc.
  Finally, the lack of an overlap in the $60-80$\,km/s gas and HESS\,J1826$-$130 would tend to rule out any direct association with SNR\,G018.6$-$0.2.

\section{Conclusions}

In this paper, we presented a detailed picture of the ISM surrounding HESS\,J1825$-$137, powered by the pulsar PSR\,J1826--1334 (here labelled P1) and discussed morphological and spectral properties of the TeV source.

Following \citet{Lem2006}'s detection of a molecular cloud  overlapping HESS\,J1826$-$130, we carried out a study of the diffuse and the dense molecular gas across 
the region using the Nanten CO(1--0) Galactic survey, the GRS $^{13}$CO(1--0) data, and our 7mm and 12mm Mopra observations tracing CS(1--0) and NH$_3$(1,1).

We observed that the bulk of the molecular gas towards HESS\,J1826$-$130 was located at \mbox{$v_{\text{lsr}}=45-60$\,km/s} which appeared consistent with the dispersion measure distance of P1.
We also noted a dense region at \mbox{(RA, Dec)=(18.421h, $-13.282^{\circ}$)} with mass \mbox{$M_{\text{H}_2}\sim1\times10^5\solmass$} and H$_2$ density \mbox{$n_{\text{H}_2}\sim7\times10^{2}$\,cm$^{-3}$} which showed enhanced turbulence.
We indicated that its CS(1--0) and $^{13}$CO(1--0) velocity structure, the presence of a UC\,H\textsc{ii} region and several OB stars and high mass star-forming regions, can be signatures of turbulent clouds caused 
by cloud cloud collisions (e.g RCW\,120 \citealt{Torii2015}).
We also suggested that its possible interaction with P1's progenitor SNR was unlikely to cause such disruptions.

The H$\alpha$  rim discovered by \citet{Stupar2008} may be associated with P1's progenitor SNR, as the distance between the H$\alpha$ rim and P1 appears consistent 
with an SNR radius $R_{\text{SNR}}\sim 120$\,pc suggested by \citet{DeJager} based on their suggestion that radius of the SNR being $\sim4$ times that of the corresponding PWN.

We found that CRs produced by the P1's progenitor SNR could explain to the TeV $\gamma$-ray emission found in HESS\,J1826$-$130.
The origin of HESS\,J1826$-$130 might also be leptonic. if associated with the PWN\,G018.5$-$0.4.
If this PWN produces the HESS\,J1826$-$130 emission, the adjacent molecular gas at \mbox{$v_{\text{lsr}}=60-80$\,km/s} may explain the TeV morphology and would suggest a PWN distance \mbox{$d=4.6$\,kpc} (near) or \mbox{$d=11.4$\,kpc} (far).
SNR\,G018.2$-$0.1 and SNR\,G018.6$-$0.2 (see \citealt{Brogan2006} and \citealt{Green2014}) are also positioned close to HESS\,J1826$-$130.
Their small angular diameters however and their offset position makes the two SNRs unlikely to be associated with HESS\,J1826$-$130.

Further VHE observations with H.E.S.S would provide more refined spatial resolution with advanced analysis (e.g \citealt{Parsons2014}).
This will enable a detail comparison between the TeV emission and the molecular gas and consequently add key information about the hadronic and/or leptonic nature of HESS\,J1826$-$130, 
and probe the diffusion of CRs into the gas (e.g see \citealt{Gabicidiff2006}).

\section*{Acknowledgments}
The Mopra radio telescope  is part of the Australia Telescope National Facility which is funded by the Australian Government for operation as a National Facility managed by CSIRO.
Operations support was provided by the University of New South Wales and the University of Adelaide.
The University of New South Wales Digital Filter Bank used for the observations with the Mopra Telescope (the UNSW–MOPS) was provided with support from the Australian Research Council (ARC).
This research has made use of the SIMBAD database, operated at CDS, Strasbourg, France.

\bibliography{paperbib}{}
\bibliographystyle{mn}

\appendix

\section[]{Beam efficiency and Coupling factor}
\label{app:couplingfactor}
The Mopra main beam only receives only a fraction of the true emission, the rest being taken by the side lobes. 
\citet{Urquhart2010} obtained the main beam efficiencies, $\eta_{\text{mb}}$, at different frequencies by calibrating the Antenna flux while observing Jupiter.
The following coefficient will be useful to get the real antenna temperature of the source: 
\begin{equation}
T_{mb}=\frac{T_{A}^{*}}{\eta_{mb}}
\end{equation}
The coupling factor fK brings the true brightness temperature for core sizes smaller than  the beam size.
Indeed,  the beam will average the signal from the core with the noise coming from the rest of the beam, minimizing its strength.
\begin{align}
\label{beamdilution1} fK  &= \left(1-\textrm{exp}\left(-\frac{4R^{2}}{\theta_{mb}^{2}}ln2\right)\right)^{-1} \\
\label{beamdilution2}\frac{\Delta\Omega_{\text{A}}}{\Delta\Omega_{\text{S}}}&=\frac{1}{fK}
\end{align}

\section[]{NH${3}$(2,2) and NH$_{3}$(3,3)}
\label{sec:appNH322}
\begin{figure*}
 \begin{minipage}{\textwidth}
   \includegraphics[width=\textwidth]{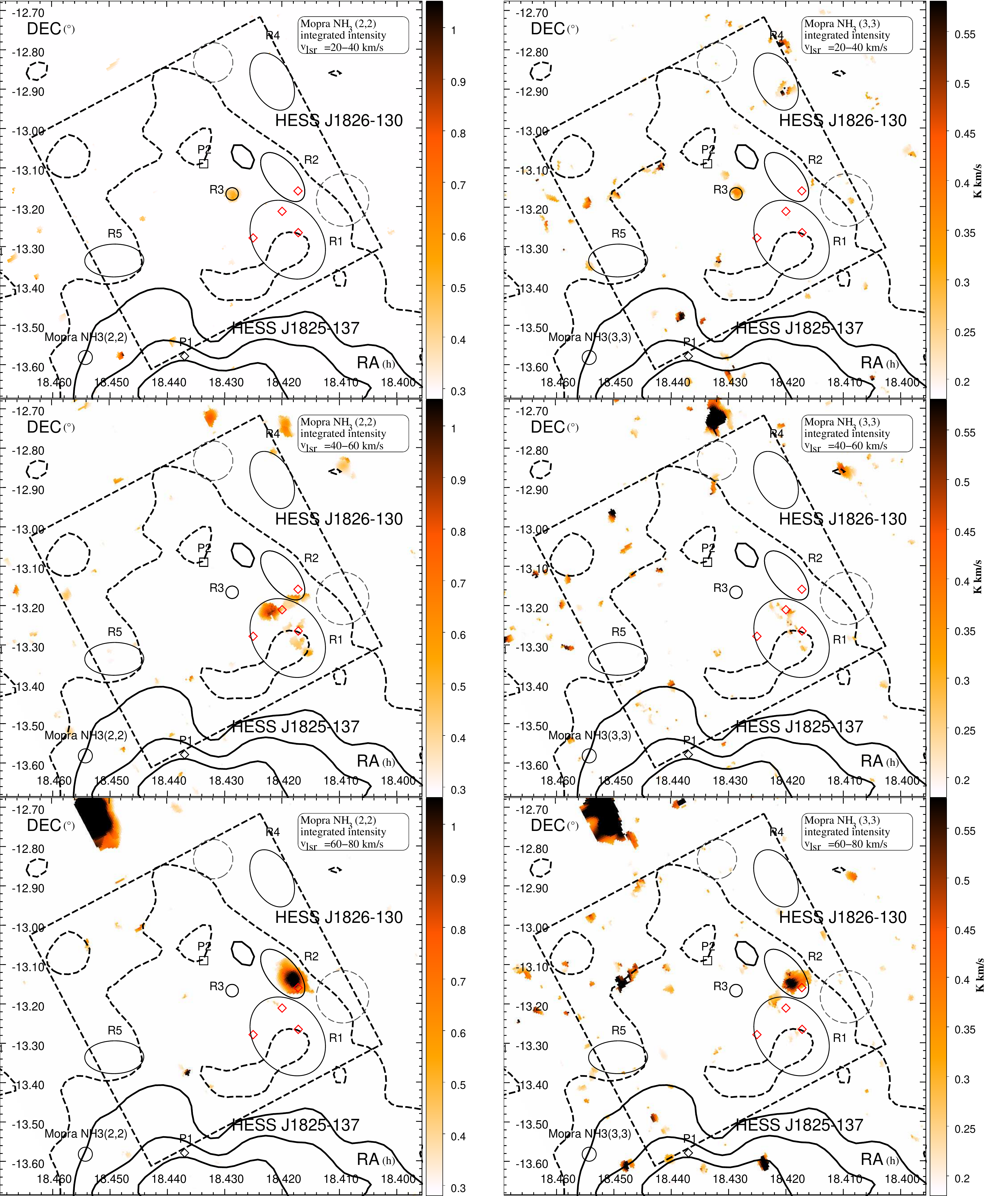}%
   \caption{NH$_3$(2,2) and NH$_3$(3,3) integrated velocity over the different kinematic velocity spans : $v_{\text{lsr}}=20-40$\,km/s, $v_{\text{lsr}}=40-60$\,km/s, $v_{\text{lsr}}=60-80$\,km/s.
   The diamonds represent the different H\textsc{ii}  regions next to \textit{R1}. The black circles represent the size of the catalogued SNRs. The region covered by our 7mm survey is shown as a black dashed rectangle.}
   \label{NH2233}
 \end{minipage}
\end{figure*}
Fig. \ref{NH2233} presents the NH$_{3}$(2,2) and NH$_{3}$(3,3) integrated intensities map for emissions located in the different kinematic velocity spans: $20-40$\,km/s, $40-60$\,km/s and $60-80$\,km/s.\\
It appears that R1, R2 and R3 shows significant emission of NH$_{3}$(2,2). However NH$_3$(3,3) are mostly found inside \textit{R2} and \textit{R3}.

\section{C$^{34}$S(1-0)}
\label{sec:34CS}
\begin{figure}
 \includegraphics[height=0.5\textwidth,angle=0]{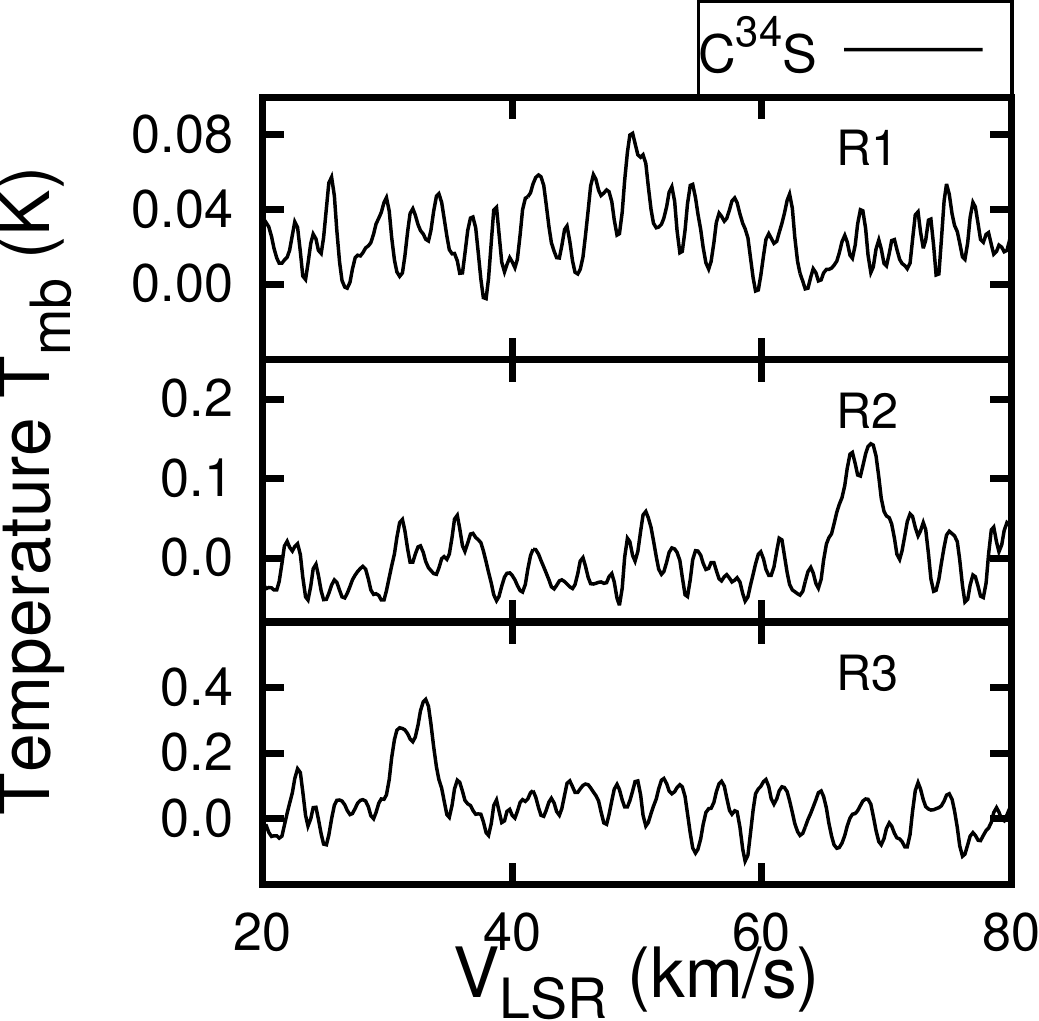}
 \caption{Averaged C$^{34}$S(1-0) emission over the 3 regions of interest \textit{R1, R2, R3}.}
 \label{34CSregions}
\end{figure}
Fig. \ref{34CSregions} displays the averaged C$^{34}$S(1-0) emission profiles found in \textit{R1, R2} and \textit{R3}.

\section[]{CO(1--0) emission at $v_{\text{lsr}}=10-25$ km/s}
\begin{figure}
\includegraphics[width=0.5\textwidth]{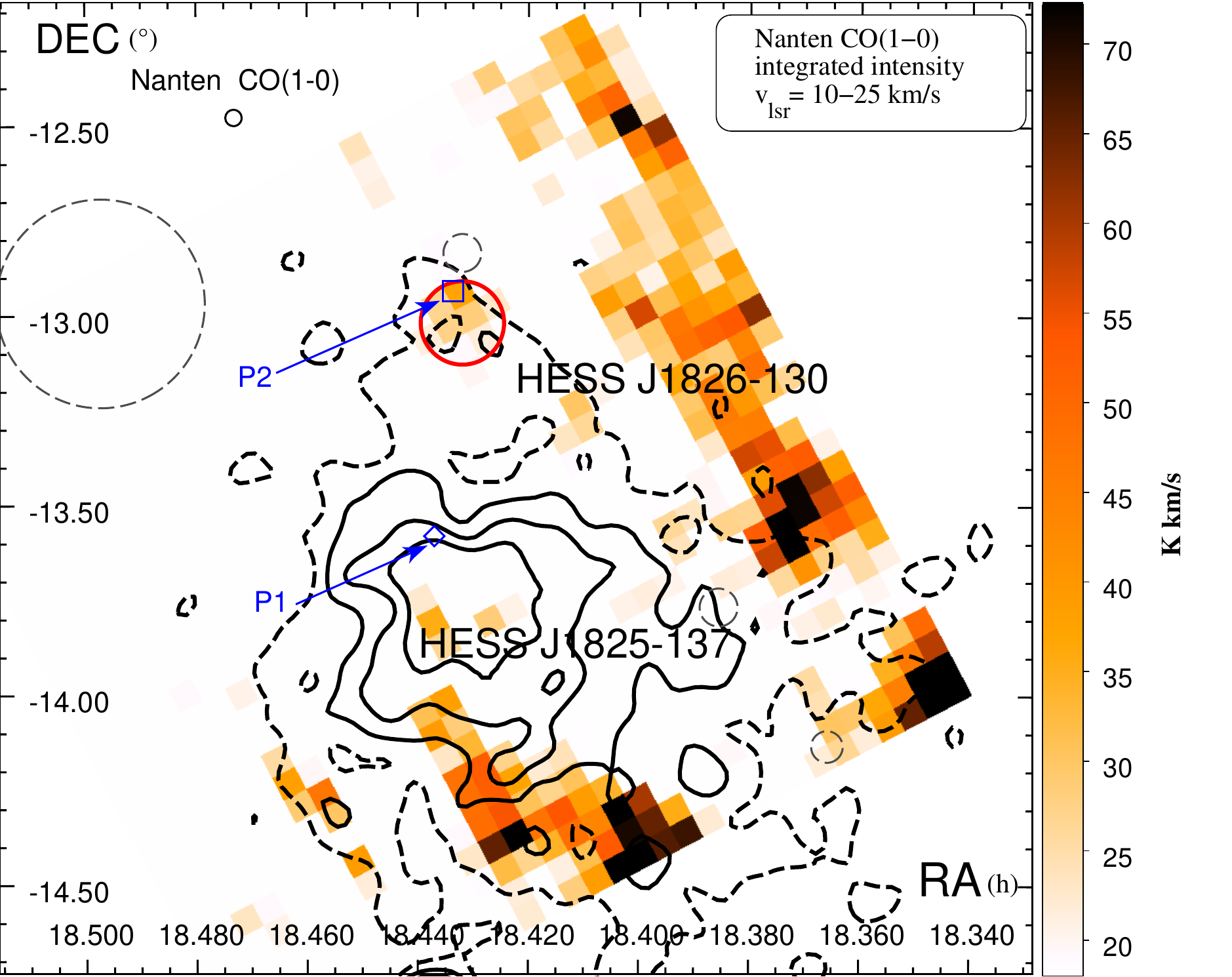}
\caption{Nanten CO(1--0) integrated intensity between $v_{\text{lsr}}=10-25$ km/s overlaid by the HESS TeV contours in black (dashed and solid). The pulsars are displayed in blue while the SNRs are represented
in grey dashed circles. The red circle highlights the molecular gas close to the pulsar PSR\,J1826--1256 (P2) highlighted in the text. The black dashed rectangle represents the region covered by our 7mm survey (see online version). }
\label{NantenCO1025}
\end{figure}

Fig.\,\ref{NantenCO1025} shows the Nanten CO(1--0) integrated intensity between $v_{\text{lsr}}=$10-25\,km/s matching the pulsar PSR\,J1826-1256 (P2) kinematic distance. The presence of a molecular cloud (red circle) spatially overlapping
 the pulsar is observed. Provided the TeV emission originated from the PWN powered by P2, the association with the molecular cloud would have significantly affected the morphology of the TeV emission.

\section[]{$^{13}$CO(1--0) emission position-velocity plots towards SNR\,G018.6$-$0.2}
\begin{figure*}
 \begin{minipage}{\textwidth}
  \includegraphics[width=\textwidth]{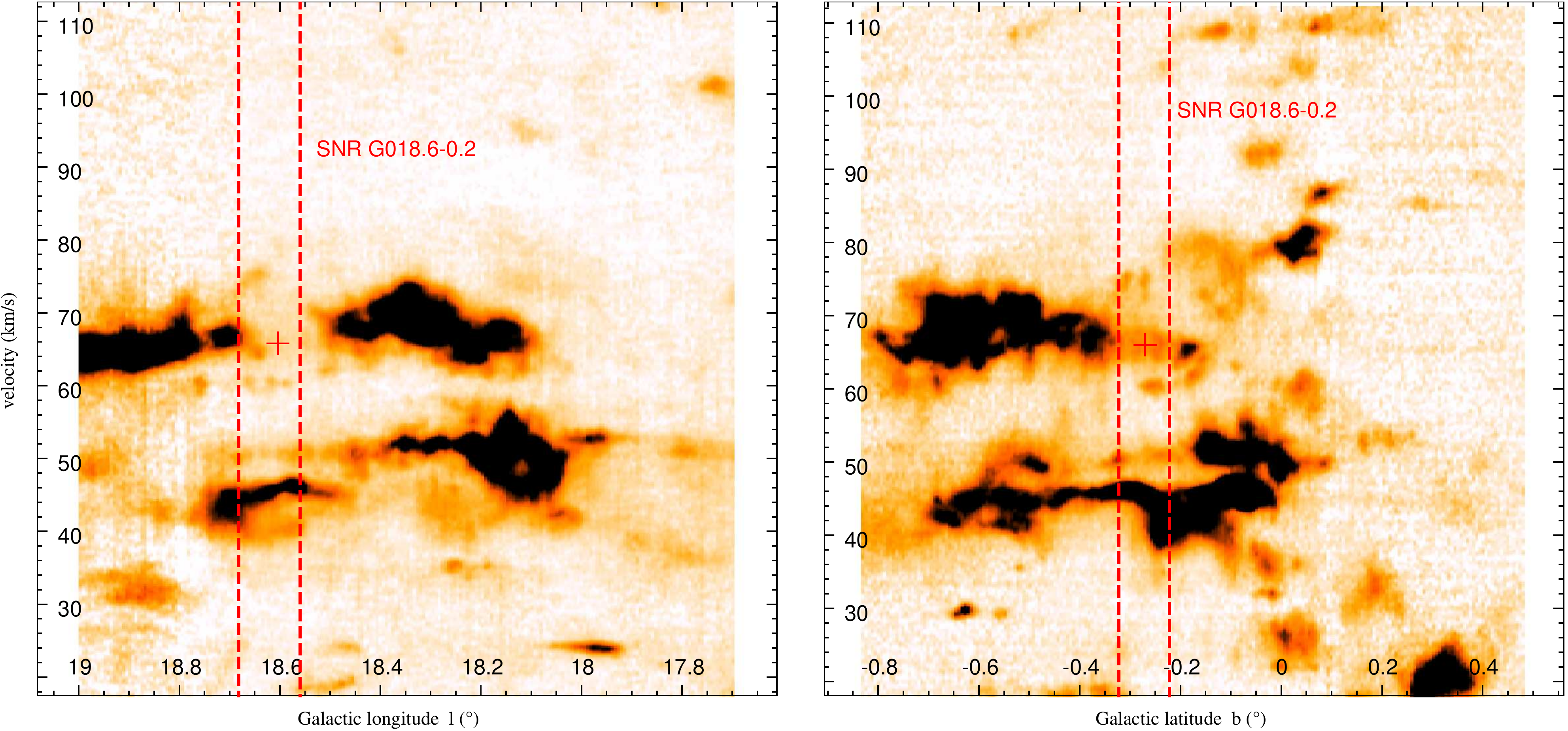}
  \caption{GRS $^{13}$CO(1--0) Galactic longitude-velocity ($l$,$v$) \textit{(left)} and Galactic latitude-velocity ($b$,$v$) \textit{(right)} maps integrated between $b=\left[-0.227 : -0.338\right]$ and $l=\left[18.554 : 18.687\right]$.
  The red cross indicates the plausible location of a weak shell spatially coincident with the SNR\,G018.6$-$0.2 whose position is delimited by the two red dashed lines (see online version for colours).}
  \label{GRSPVplot}
 \end{minipage}
\end{figure*}

Fig.\,\ref{GRSPVplot} shows the $^{13}$CO(1--0) position-velocity (in galactic coordinates) map towards the SNR\,G018.6$-$0.2.
We observe from the Galactic longitude-velocity plot a drop of ${13}$CO at $v_{\text{lsr}}=60-70$ km/s towards the SNR (whose boundaries are shown as red dashed lines).
However, weak emission appears at $v_{\text{lsr}}\sim60$\,km/s and 75\,km/s (appearing in both maps) suggesting that gas may have been accelerated.
Its spatial coincidence with the SNR position support the scenario of a putative shell produced by the SNR progenitor star.
\label{app:GRSPVplot}

\section[]{HC$_{3}$N(5-4,F=4-3)}
\label{app:HCCCN}

\begin{figure*}
\begin{minipage}{\textwidth}
 \includegraphics[width=\textwidth]{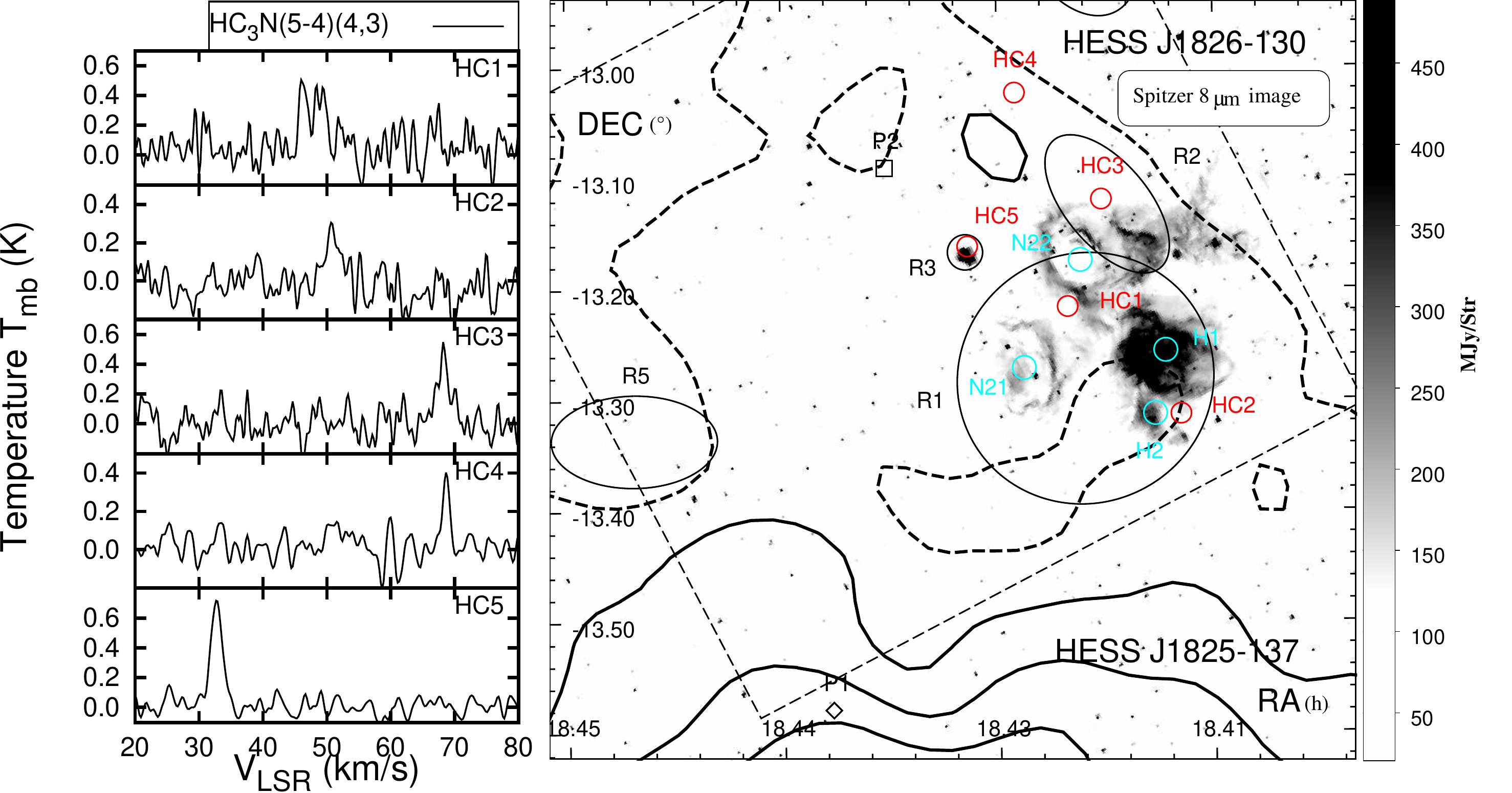}
 \caption{Spitzer 8$\mu$m map towards HESS\,J1826$-$130. The red circles indicate the regions where HC$_3$N(5--4,F=4--3) emission are found.
  Their respective spectra are shown in the left hand side.
   The aforementioned H\textsc{ii} regions are displayed as cyan circles (see online version for colours).}
 \label{spitzerhCCN}
 \end{minipage}
\end{figure*}

Fig.\,\ref{spitzerhCCN} shows the position of the observed HC$_3$N(5-4,F=4-3) position, and their respective spectra.

\section{Physical parameters derivation}
\subsection{CS parameters}
\textbf{CS(1--0) optical depth}:
\begin{equation}
\label{tauCS}\frac{W_{CS(1-0)}}{W_{C^{34}S(1-0)}} = \frac{1-e^{-\tau_{CS(1-0)}}}{1-e^{\alpha\tau_{CS}(1-0)}}  
\end{equation}
where $\alpha$ represents the relative abundance ratio C$^{34}$S/CS=0.04 (see section 4).\newline\newline
\textbf{Column density of energy state 1} :
\begin{equation}
 \label{CSeq2} N_{CS_1} = \frac{8k \pi\nu_{10}^{2}}{A_{10}hc^{3}}\left(\frac{\Delta\Omega_{\text{A}}}{\Delta\Omega_{\text{S}}}\right)\left(\frac{\tau_{\text{CS(1-0)}}}{1-e^{-\tau_{\text{CS(1-0)}}}}\right)\int{T_{\text{mb}}\left(v\right)dv} 
\end{equation}
where $\nu_{10}$ and $A_{10}$ is the rest frequency and Einstein's coefficient of the CS(1--0) emission respectively. $\Delta\Omega_\text{A}$ and $\Delta\Omega_\text{S}$ are the antenna and source solid angle respectively.
In the case where $\Delta\Omega_\text{A}>\Delta\Omega_\text{S}$, we use Eqs.\,\ref{beamdilution1}\,and\,\ref{beamdilution2}.\newline\newline
\textbf{CS column density} :
\begin{equation}
 \label{CSeq3} N_{CS} = N_{CS_1}\left(1+\frac{1}{3}e^{2.35/T_{\text{kin}}}+\frac{5}{3}e^{-4.7/T_{\text{kin}}}+\text{...}\right) 
\end{equation}

\subsection{NH$_3$ parameters}
\textbf{NH$_3$(1,1) optical depth} :
\begin{eqnarray}
\label{tauNH3} \frac{T^{*}_{\text{A}_{\text{m}}}}{T^{*}_{\text{A}_{\text{s}}}} & = & \frac{1-e^{-\tau_{\text{m}}}}{1-e^{\alpha\tau_{\text{m}}}} \text{ for NH}_3\\
\label{tauNH3b} \tau_{\text{NH}_3\left(1,1\right)} &= &  \frac{\tau_{\text{m}}}{0.5}
\end{eqnarray}
Now,  $\tau_\text{m}$ is the optical depth of the main emission and $\alpha$ represents the relative strength of the satellite line compared to the main line.\newline\newline
\textbf{NH$_3$(1,1) and NH$_3$(2,2) column densities} :
\begin{align}
\label{NH3eq3} N_{1,1} & = \frac{8k \pi\nu_{11}^{2}}{A_{11}hc^{3}}\left(\frac{\Delta\Omega_{\text{A}}}{\Delta\Omega_{\text{S}}}\right)\left(\frac{\tau_{\text{NH}_3\text{(1,1)}}}{1-e^{-\tau_{\text{NH}_3\text{(1,1)}}}}\right)\int{T_{\text{mb}}\left(v\right)dv} \\ 
\label{NH3eq4} N_{2,2} & = \frac{8k \pi\nu_{22}^{2}}{A_{22}hc^{3}}\left(\frac{\Delta\Omega_{\text{A}}}{\Delta\Omega_{\text{S}}}\right)\left(\frac{\tau_{\text{NH}_3\text{(2,2)}}}{1-e^{-\tau_{\text{NH}_3\text{(2,2)}}}}\right)\int{T_{\text{mb}}\left(v\right)dv} 
\end{align}
where $\nu_{11}$ and $\nu_{22}$ are the rest frequencies of the NH$_3$(1,1) and NH$_3$(2,2) emission respectively. $\Delta\Omega_\text{A}$ and $\Delta\Omega_\text{S}$ are the antenna and source solid angle respectively.
In the case where $\Delta\Omega_\text{A}>\Delta\Omega_\text{S}$, we use Eqs.\,\ref{beamdilution1}\,and\,\ref{beamdilution2}.\newline\newline
\textbf{Rotational and kinetic temperature}
\begin{align}
\label{NH3eq6} T_{\text{rot}} & = -\frac{41}{ln\left(3N_{2,2}/\left(5N_{1,1}\right)\right)} \\ 
\label{NH3eq7} T_{\text{kin}} & = T_{\text{rot}}\left(1-\frac{T_{\text{rot}}}{42}\log\left(1+1.1\exp\left(-16/T_{\text{rot}}\right)\right)\right)^{-1} 
\end{align}
\newline\newline
\textbf{NH$_3$ column density}
\begin{equation}
\label{NH3eq5} N_{tot} = N_{1,1}\left(1+\frac{1}{3}e^{23.26/T_{\text{kin}}}+\frac{14}{3}e^{-100.25/T_{\text{kin}}}+\text{...}\right) 
\end{equation}
\subsection{Mass and density}
\begin{align}
\label{H2mass} M_{\text{H}_2}\left(X\right)=\mu m_{\text{H}}\pi abN_{\text{H}_2}\\
\label{H2dens} n_{\text{H}_2}\left(X\right)=\frac{M}{4/3\left(\mu m_{\text{H}}\right)\pi ab^2}
\end{align}
  
where $\mu=2.8$ represents the weight factor assuming  the molecular cloud  consisted of 20\% of He, $X$ was either NH$_3$ or CS and $a$, $b$ was the semi-minor  and semi-major axis respectively.
$ab^2$ assumes the cloud also extends at distance $b$ in the z-direction.

\section[]{Gas parameters derived for regions \textit{R1} to \textit{R6}}
\begin{table*}
\centering
\begin{minipage}{\textwidth}
\begin{subtables}\label{MASStot}
\centering
\caption{Parameters derived towards region \textit{R1} using the molecular transition CS(1-0), CO(1-0), $^{13}$CO(1-0), NH$_{3}$(1,1), NH$_3$(2,2) and C$^{34}$S(1-0) when emission are found.
    The labels \textit{a},\textit{b},\textit{c},\textit{d} denote a distinct emission. Lower limits have been derived when we assumed $\tau=0$ }
\begin{tabular}{ccccccccc}
    \toprule
 CS &R1 & distance & $\frac{W_{\text{CS}\left(1-0\right)}}{W_{\text{C}^{34}\text{S}\left(1-0\right)}}$  & $\tau_{\text{CS}\left(1-0\right)}$  & $N_{\text{CS}}$ [10$^{12}$]$^{a}$ & $N_{\text{H}_2}$[10$^{20}$]$^{ab}$&  $M_{\text{H}_2}^{abc}$ & $n_{\text{H}_2}^{abc}$\\
    &  & (kpc) & & & (cm$^{-2}$)& (cm$^{-2}$)& ($\solmass$)& (cm$^{-3}$) \\
    \midrule
    
    & \multirow{2}{*}{R1a} & near : 3.9 & \multirow{2}{*}{23.5}  & \multirow{2}{*}{0.1} & \multirow{2}{*}{80} &\multirow{2}{*}{200} &  $1.0\times10^{5}$ & $7.5\times10^{2}$\\
    & & far : 12.2 & & & & &  $1.0\times10^{6}$ & $3.8\times10^{2}$ \\
    \\
    & \multirow{2}{*}{R1c} & near : 3.5 & \multirow{2}{*}{-} & \multirow{2}{*}{-} & \multirow{2}{*}{$>2$} &\multirow{2}{*}{$>5$} & $>1.8\times10^{3}$ & $>1.4\times10^{1}$ \\
    & & far : 12.6 & & & & & $>2.4\times10^{4}$ & $>4$ \\
    \\
    & \multirow{2}{*}{R1e} & near : 4.7 & \multirow{2}{*}{-}  & \multirow{2}{*}{-} & \multirow{2}{*}{$>13$} &\multirow{2}{*}{$>31$} & $>2.4\times10^{4}$ & $>7.4\times10^{1}$ \\
    & & far : 11.4 & & & & & $>1.4\times10^{5}$ & $>3.0\times10^{1}$ \\
    \bottomrule
    \end{tabular}
    \vspace{0.5cm}
    \centering
  \begin{tabular}{ccccccccc}
  \toprule
NH$_3$ &R1 & distance & $\tau_{\text{NH}_3\left(1,1\right)}$& $T_{\text{kin}}$ & $N_{\text{NH}_3}$ [10$^{12}$] & $N_{\text{H}_2}\left[10^{20}\right]^{ab}$ & $M_{\text{H}_2}^{abc}$ & $n_{\text{H}_2}^{abc}$ \\
& & (kpc)& & (K)& (cm$^{-2}$)& (cm$^{-2}$)& ($\solmass$)& (cm$^{-3}$)\\
\midrule
 &\multirow{2}{*}{R1a}& near : 3.9 & \multirow{2}{*}{0.43} & \multirow{2}{*}{18} & \multirow{2}{*}{46} & \multirow{2}{*}{46} & $1.4\times10^{4}$ & $2.0\times10^{2}$ \\
& & far : 12.2 &  & & & & $9.7\times10^{4}$ & $6.3\times10^{1}$ \\
\\
& \multirow{2}{*}{R1e}& near : 4.7 & \multirow{2}{*}{-} & \multirow{2}{*}{-} & \multirow{2}{*}{$>14$} & \multirow{2}{*}{$>14$} & $>6.1\times10^{3}$ & $>4.8\times10^{1}$\\
& & far : 11.4 &  & & & & $>3.6\times10^{4}$& $>2.0\times10^{1}$\\
\bottomrule
  \end{tabular}
  \vspace{0.5cm}
  \centering
    \begin{tabular}{ccccccc}
    \cmidrule[1pt]{1-6}
  CO & R1 & distance & $M_{\text{H}_2}^{d}$ & $\Mvir\left(^{13}\text{CO}\right)^{e}$& $n_{\text{H}_2}^{e}$ \\
  & & (kpc)   &($\solmass$) & ($\solmass$) & (cm$^{-3}$) \\
  \cmidrule{1-6}
    &\multirow{2}{*}{R1a}& near : 3.9  & $1.2\times10^{5}$& $4.8\times10^{4}-2.1\times10^{5}$& $9.6\times10^{2}$ \\
    & & far : 12.2 & $1.1\times10^{6}$ & $1.5\times10^{5}-5.3\times10^{5}$& $3.0\times10^{2}$ \\
    \\
     &\multirow{2}{*}{R1c}& near : 3.5  & $2.0\times10^{4}$& $6\times10^{3}-2.1\times10^{4}$& $2.2\times10^{2}$ \\
    & & far : 12.6 & $2.6\times10^{5}$&$2.1\times10^{4}-7.6\times10^{4}$ & $6.0\times10^{1}$ \\
    \\
     &\multirow{2}{*}{R1d}& near : 4.3 & $6.8\times10^{3}$& $7.0\times10^{3}-2.6\times10^{4}$& $4.3\times10^{1}$ \\
    & & far : 11.8 & $5.2\times10{4}$&$2.0\times10^{4}-7.1\times10^{4}$& $1.5\times10^{1}$ \\
    \\
     &\multirow{2}{*}{R1e}& near : 4.7 & $6.8\times10^{4}$& $1.5\times10^{5}-5.3\times10^{5}$& $3.2\times10^{2}$ \\
    & & far : 12.2 & $4.2\times10^{5}$& $7.8\times10^{4}-2.8\times10^{5}$ & $1.3\times10^{2}$\\
    \\
    \cmidrule[1pt]{1-6}
    \multicolumn{7}{p{12cm}}{\footnotesize{$^{a}$:Parameters have been derived using the LTE assumption.}}\\
    \multicolumn{7}{p{12cm}}{\footnotesize{$^{b}$:The H$_{2}$ physical parameters derived using a NH$_3$ abundance ratio $\chi_{\text{NH}_3}=1\times10^{-8}$ and using a CS abundance ratio $\chi_{\text{CS}}=4\times10^{-9}$. }}\\
    \multicolumn{7}{p{12cm}}{\footnotesize{$^{c}$:H$_{2}$ mass and density from CS and NH$_3$ have been computed assuming the observed region is spherical or ellipsoid and whose size are given in table \ref{fitlabel} and table \ref{fitlabelbis}. }}\\
    \multicolumn{7}{p{12cm}}{\footnotesize{$^{d}$:H$_{2}$ mass are derived using a X$_{\text{CO}}=2.0\times10^{20}$cm$^{-2}$ (K km/s)$^{-1}$ and assuming a spherical region.}}\\
    \multicolumn{7}{p{12cm}}{\footnotesize{$^{e}$:Virial mass is computed using CO(1-0) and $^{13}$CO(1-0) emission FWHM and assuming a spherical region. The left value represents the Virial mass for a 1/$r^2$ density distribution whereas the right value indicate the value for a Gaussian distribution.}}\\
    \multicolumn{7}{p{12cm}}{\footnotesize{.}}
    
    \end{tabular}
    
\end{subtables}

\end{minipage}

\end{table*}

\setcounter{table}{0}
\begin{table*}
\centering
\begin{minipage}{\textwidth}
\begin{subtables}
\setcounter{table}{1}
\footnotesize
\centering
\caption{Parameters derived towards region \textit{R2} using the molecular transition CS(1-0), CO(1-0), $^{13}$CO(1-0), NH$_{3}$(1,1), NH$_3$(2,2) and C$^{34}$S(1-0) when emission are found.
    The labels \textit{a},\textit{b},\textit{c},\textit{d} denote a distinct emission. Lower limits have been derived when we assumed $\tau=0$ }
\begin{tabular}{ccccccccc}
    \toprule
 CS &R2 & distance & $\frac{W_{\text{CS}\left(1-0\right)}}{W_{\text{C}^{34}\text{S}\left(1-0\right)}}$ & $\tau_{\text{CS}\left(1-0\right)}$  & $N_{\text{CS}}$ [10$^{12}$]$^{a}$ & $N_{\text{H}_2}$[10$^{20}$]$^{ab}$&  $M_{\text{H}_2}^{abc}$ & $n_{\text{H}_2}^{abc}$ \\
    &  & (kpc) & & & (cm$^{-2}$)& (cm$^{-2}$)& ($\solmass$)& (cm$^{-3}$) \\
    \midrule
    & \multirow{2}{*}{R2a} & near : 4.0 & \multirow{2}{*}{-} & \multirow{2}{*}{-} & \multirow{2}{*}{$>14$} &\multirow{2}{*}{$>36$} & $>3.4\times10^{3}$ & $>3.3\times10^{2}$  \\
    & & far : 12.1 & & & & & $>3.2\times10^{4}$ & $>1.1\times10^{2}$ \\
    \\
     & \multirow{2}{*}{R2e} & near : 4.7 & \multirow{2}{*}{13.3} & \multirow{2}{*}{1.5} & \multirow{2}{*}{100} &\multirow{2}{*}{250} & $3.4\times10^{4}$ & $2.0\times10^{3}$ \\
    & & far : 11.4 & & & & & $2.0\times10^{5}$ & $8.5\times10^{2}$ \\
    \bottomrule
    \end{tabular}
\vspace{0.5cm}
\centering
  \begin{tabular}{ccccccccc}
  \toprule
NH$_3$ &R2 & distance & $\tau_{\text{NH}_3\left(1,1\right)}$& $T_{\text{kin}}$ & $N_{\text{NH}_3}$ [10$^{12}$] & $N_{\text{H}_2}\left[10^{20}\right]^{ab}$ & $M_{\text{H}_2}^{abc}$ & $n_{\text{H}_2}^{abc}$  \\
& & (kpc)& & (K)& (cm$^{-2}$)& (cm$^{-2}$)& ($\solmass$)& (cm$^{-3}$)\\
\midrule
& \multirow{2}{*}{R2a}& near : 4.0 & \multirow{2}{*}{-} & \multirow{2}{*}{-} & \multirow{2}{*}{$>5$} & \multirow{2}{*}{$>5$} & $>4.5\times10^{2}$ & $>4.3\times10^{1}$  \\
& & far : 12.1 &  & & & & $>4.0\times10^{3}$& $>1.4\times10^{1}$ \\
\\
 &\multirow{2}{*}{R2e}& near : 4.7 & \multirow{2}{*}{3.5} & \multirow{2}{*}{11} & \multirow{2}{*}{600} & \multirow{2}{*}{600} & $8.1\times10^{4}$ & $4.8\times10^{3}$ \\
& & far : 11.4 &  & & & & $4.8\times10^{5}$& $2.0\times10^{3}$\\
\midrule
  \end{tabular}
  \vspace{0.5cm}
  \centering

    \begin{tabular}{ccccccc}
    \cmidrule[1pt]{1-6}
   CO & R2 & distance & $M_{\text{H}_2}^{d}$ & $\Mvir\left(^{13}\text{CO}\right)^{e}$& $n_{\text{H}_2}^{e}$ \\
  & & (kpc)  &($\solmass$) & ($\solmass$)& (cm$^{-3}$) \\
  \cmidrule{1-6}
   &\multirow{2}{*}{R2a}& near : 4.0 & $1.3\times10^{4}$& $4.2\times10^{3}-2.5\times10^{4}$& $1.2\times10^{3}$ \\
    & & far: 12.1 & $1.1\times10^{5}$& $1.3\times10^{4}-4.4\times10^{4}$& $8.9\times10^{2}$ \\
    \\
     &\multirow{2}{*}{R2c}& near : 3.5 & $8.8\times10^{3}$&$8.9\times10^{4}-3.1\times10^{5}$& $3.9\times10^{2}$ \\
    & & far: 12.6 & $1.1\times10^{5}$& $3.2\times10^{5}-1.1\times10^{6}$& $2.5\times10^{2}$ \\
    \\
    &\multirow{2}{*}{R2e}& near : 4.7 & $2.9\times10^{4}$ & $3.1\times10^{4}-1.1\times10^{5}$& $1.7\times10^{3}$ \\
    & & far: 11.4 & $2.0\times10^{5}$& $7.5\times10^{4}-2.7\times10^{5}$&$6.4\times10^{2}$ \\
    \cmidrule[1pt]{1-6}
    \multicolumn{7}{p{12cm}}{\footnotesize{$^{a}$:Parameters have been derived using the LTE assumption.}}\\
    \multicolumn{7}{p{12cm}}{\footnotesize{$^{b}$:The H$_{2}$ physical parameters derived using a NH$_3$ abundance ratio $\chi_{\text{NH}_3}=1\times10^{-8}$ and using a CS abundance ratio $\chi_{\text{CS}}=4\times10^{-9}$. }}\\
    \multicolumn{7}{p{12cm}}{\footnotesize{$^{c}$:H$_{2}$ mass and density from CS and NH$_3$ have been computed assuming the observed region is spherical or ellipsoid and whose size are given in table \ref{fitlabel} and table \ref{fitlabelbis}. }}\\
    \multicolumn{7}{p{12cm}}{\footnotesize{$^{d}$:H$_{2}$ mass are derived using a X$_{\text{CO}}=2.0\times10^{20}$cm$^{-2}$ (K km/s)$^{-1}$ and assuming a spherical region.}}\\
    \multicolumn{7}{p{12cm}}{\footnotesize{$^{e}$:Virial mass is computed using CO(1-0) and $^{13}$CO(1-0) emission FWHM and assuming a spherical region. The left value represents the Virial mass for a 1/$r^2$ density distribution whereas the right value indicate the value for a Gaussian distribution.}}\\
    \end{tabular}

\end{subtables}
\end{minipage}
\end{table*}

\clearpage
\setcounter{table}{0}
\begin{table*}
\centering
\begin{minipage}{\textwidth}
\begin{subtables}
\setcounter{table}{2}
\footnotesize
\centering
\caption{Parameters derived towards region \textit{R3} using the molecular transition CS(1-0), CO(1-0), $^{13}$CO(1-0), NH$_{3}$(1,1), NH$_3$(2,2) and C$^{34}$S(1-0) when emission are found.
    The labels \textit{a},\textit{b},\textit{c} denote a distinct emission. Lower limits have been derived when we assumed $\tau=0$ }

\begin{tabular}{ccccccccc}
    \toprule
 CS &R3 & distance & $\frac{W_{\text{CS}\left(1-0\right)}}{W_{\text{C}^{34}\text{S}\left(1-0\right)}}$ & $\tau_{\text{CS}\left(1-0\right)}$  & $N_{\text{CS}}$ [10$^{12}$]$^{a}$ & $N_{\text{H}_2}$[10$^{20}$]$^{ab}$&  $M_{\text{H}_2}^{abc}$ & $n_{\text{H}_2}^{abc}$\\
    &  & (kpc) & &  & (cm$^{-2}$)& (cm$^{-2}$)& ($\solmass$)& (cm$^{-3}$) \\
    \midrule
    & \multirow{2}{*}{R3e} & near : 4.8 & \multirow{2}{*}{-}  & \multirow{2}{*}{-} & \multirow{2}{*}{$>7$} &\multirow{2}{*}{$>16$} & $>2.5\times10^{2}$ & $>2.0\times10^{2}$ \\
    & & far : 11.3 &  & & & & $>1.7\times10^{3}$& $>1.1\times10^{2}$ \\
    \\
    & \multirow{2}{*}{R3f} & near : 2.9 & \multirow{2}{*}{8.3} & \multirow{2}{*}{3.0} & \multirow{2}{*}{520} &\multirow{2}{*}{1300} & $7.4\times10^{3}$ & $3.5\times10^{4}$\\
    & & far : 13.3 & & & & & $1.5\times10^{5}$ &  $7.6\times10^{3}$ \\

    \bottomrule
    \end{tabular}
\vspace{0.5cm}
\centering
  \begin{tabular}{ccccccccc}
  \toprule
NH$_3$ &R3 & distance & $\tau_{\text{NH}_3\left(1,1\right)}$& $T_{\text{kin}}$ & $N_{\text{NH}_3}$ [10$^{12}$] & $N_{\text{H}_2}\left[10^{20}\right]^{ab}$ & $M_{\text{H}_2}^{abc}$ & $n_{\text{H}_2}^{abc}$ \\
& & (kpc)& & (K)& (cm$^{-2}$)& (cm$^{-2}$)& ($\solmass$)& (cm$^{-3}$)   \\
\midrule
 &\multirow{2}{*}{R3f}& near : 2.9 & \multirow{2}{*}{2.0} & \multirow{2}{*}{20} & \multirow{2}{*}{165} & \multirow{2}{*}{165} & $9.5\times10^{2}$ & $4.4\times10^{3}$ \\
& & far : 13.3 &  & & & & $1.3\times10^{4}$& $6.2\times10^{2}$\\
\bottomrule
  \end{tabular}

    \centering
    \vspace{0.5cm}
    \begin{tabular}{ccccccc}
    \cmidrule[1pt]{1-6}
   CO & R3 & distance & $M_{\text{H}_2}^{d}$ & $\Mvir\left(^{13}\text{CO}\right)^{e}$ &$n_{\text{H}_2}^{e}$ \\
  & & (kpc)  &($\solmass$) & ($\solmass$) & (cm$^{-3}$) \\
  \cmidrule{1-6}
    &\multirow{2}{*}{R3a}& near : 2.9  & $3.3\times10^{3}$&$1.2\times10^{3}-4.4\times10^{3}$& $1.6\times10^{4}$ \\
    & & far : 13.3 & $7.2\times10^{4}$& $5.7\times10^{3}-2.0\times10^{4}$& $4.3\times10^{3}$ \\
    \\
     &\multirow{2}{*}{R3e}& near : 4.8  & $2.3\times10^{3}$ & $2.7\times10^{3}-9.6\times10^{3}$& $2.5\times10^{3}$  \\
    & & far : 11.3 & $1.3\times10^{4}$ & $6.4\times10^{4}-2.3\times10^{5}$& $1.1\times10^{3}$ \\
    \\
     &\multirow{2}{*}{R3f}& near : 3.8 &$1.4\times10^{3}$ & $2.1\times10^{3}-7.6\times10^{3}$& $3.1\times10^{3}$ \\
    & & far : 12.3 & $1.5\times10^{4}$ & $6.9\times10^{3}-2.5\times10^{4}$&$9.3\times10^{2}$ \\
    \cmidrule[1pt]{1-6}
    \multicolumn{7}{p{12cm}}{\footnotesize{$^{a}$:Parameters have been derived using the LTE assumption.}}\\
    \multicolumn{7}{p{12cm}}{\footnotesize{$^{b}$:The H$_{2}$ physical parameters derived using a NH$_3$ abundance ratio $\chi_{\text{NH}_3}=1\times10^{-8}$ and using a CS abundance ratio $\chi_{\text{CS}}=4\times10^{-9}$. }}\\
    \multicolumn{7}{p{12cm}}{\footnotesize{$^{c}$:H$_{2}$ mass and density from CS and NH$_3$ have been computed assuming the observed region is spherical or ellipsoid and whose size are given in table \ref{fitlabel} and table \ref{fitlabelbis}. }}\\
    \multicolumn{7}{p{12cm}}{\footnotesize{$^{d}$:H$_{2}$ mass are derived using a X$_{\text{CO}}=2.0\times10^{20}$cm$^{-2}$ (K km/s)$^{-1}$ and assuming a spherical region.}}\\
    \multicolumn{7}{p{12cm}}{\footnotesize{$^{e}$:Virial mass is computed using CO(1-0) and $^{13}$CO(1-0) emission FWHM and assuming a spherical region. The left value represents the Virial mass for a 1/$r^2$ density distribution whereas the right value indicate the value for a Gaussian distribution.}}\\
    \end{tabular}

\end{subtables}
\end{minipage}
\end{table*}


\clearpage
\setcounter{table}{0}
\begin{table*}
\centering
\begin{minipage}{\textwidth}
\begin{subtables}
\setcounter{table}{3}
\footnotesize
\centering
\caption{Parameters derived towards region \textit{R4} using the molecular transition CS(1-0), CO(1-0), $^{13}$CO(1-0), NH$_{3}$(1,1), NH$_3$(2,2) and C$^{34}$S(1-0) when emission are found.
    The labels \textit{a},\textit{b},\textit{c} denote a distinct emission. Lower limits have been derived when we assumed $\tau=0$ }
 \begin{tabular}{ccccccccc}
    \toprule
 CS &R4 & distance & $\frac{W_{\text{CS}\left(1-0\right)}}{W_{\text{C}^{34}\text{S}\left(1-0\right)}}$ & $\tau_{\text{CS}\left(1-0\right)}$  & $N_{\text{CS}}$ [10$^{12}$]$^{a}$ & $N_{\text{H}_2}$[10$^{20}$]$^{ab}$&  $M_{\text{H}_2}^{abc}$ & $n_{\text{H}_2}^{abc}$ \\
    &  & (kpc) & & & (cm$^{-2}$)& (cm$^{-2}$)& ($\solmass$)& (cm$^{-3}$) \\
    \midrule
    
    & \multirow{2}{*}{R4a} & near : 4.0 & \multirow{2}{*}{-} & \multirow{2}{*}{-} & \multirow{2}{*}{$>23$} &\multirow{2}{*}{$>58$} & $>7.4\times10^{3}$ & $>4.1\times10^{2}$ \\
    & & far : 12.2 & &  & & & $>7.0\times10^{4}$ & $>1.4\times10^{2}$ \\
    \\
    & \multirow{2}{*}{R4b} & near : 3.6 & \multirow{2}{*}{-} & \multirow{2}{*}{-} & \multirow{2}{*}{$>15$} &\multirow{2}{*}{$>38$} &$>5.0\times10^{3}$  & $>2.7\times10^{2}$\\
    & & far : 12.5 & & & & & $>4.9\times10^{4}$ & $>9.0\times10^{1}$ \\
    
    \bottomrule
    \end{tabular}
\vspace{0.5cm}
  \begin{tabular}{ccccccccc}
  \toprule
NH$_3$ &R4 & distance & $\tau_{\text{NH}_3\left(1,1\right)}$& $T_{\text{kin}}$ & $N_{\text{NH}_3}$ [10$^{12}$] & $N_{\text{H}_2}\left[10^{20}\right]^{ab}$ & $M_{\text{H}_2}^{abc}$ & $n_{\text{H}_2}^{abc}$ \\
& & (kpc)& & (K)& (cm$^{-2}$)& (cm$^{-2}$)& ($\solmass$)& (cm$^{-3}$)\\
\midrule
 &\multirow{2}{*}{R4a}& near : 4.0 & \multirow{2}{*}{-} & \multirow{2}{*}{-} & \multirow{2}{*}{$>19$} & \multirow{2}{*}{$>19$} & $>2.0\times10^{3}$ & $>1.5\times10^{2}$ \\
& & far : 12.2 &  & & & & $>2.4\times10^{4}$& $>4.4\times10^{1}$\\

&\multirow{2}{*}{R4b}& near : 3.6 & \multirow{2}{*}{-} & \multirow{2}{*}{-} & \multirow{2}{*}{$>10$} & \multirow{2}{*}{$>10$} & $>1.2\times10^{3}$ & $>7.0\times10^{1}$ \\
& & far : 12.5 &  & & & & $>1.1\times10^{4}$& $>2.3\times10^{1}$ \\
\bottomrule
  \end{tabular}
  
  \vspace{0.5cm}

    \begin{tabular}{ccccccc}
    \cmidrule[1pt]{1-6}
   CO & R4 & distance & $M_{\text{H}_2}^{d}$& $\Mvir\left(^{13}\text{CO}\right)^{e}$& $n_{\text{H}_2}^{e}$ \\
  & & (kpc)  &($\solmass$) & ($\solmass$) & (cm$^{-3}$) \\
  \cmidrule{1-7}
    &\multirow{2}{*}{R4a}& near : 4.0 & $1.6\times10^{4}$& $7.2\times10^{3}-2.5\times10^{4}$&$9.3\times10^{2}$ \\
    & & far : 12.2 & $1.5\times10^{5}$&$2.2\times10^{4}-7.8\times10^{4}$&$3.0\times10^{2}$ \\
    \\
     &\multirow{2}{*}{R4b}& near : 3.6 & $1.1\times10^{4}$& $6.1\times10^{3}-2.2\times10^{4}$&$8.2\times10^{2}$ \\
    & & far : 12.5  & $1.3\times10^{5}$ & $2.1\times10^{4}-7.5\times10^{4}$&$2.5\times10^{2}$\\
    \\
     &\multirow{2}{*}{R4e}& near : 4.6 & $9.0\times10^{3}$& $1.9\times10^{4}-6.7\times10^{4}$&$3.6\times10^{2}$ \\
    & & far : 11.5 & $5.6\times10^{4}$& $4.7\times10^{4}-1.7\times10^{5}$ & $1.1\times10^{2}$  \\
    \cmidrule[1pt]{1-6}
    \multicolumn{7}{p{12cm}}{\footnotesize{$^{a}$:Parameters have been derived using the LTE assumption.}}\\
    \multicolumn{7}{p{12cm}}{\footnotesize{$^{b}$:The H$_{2}$ physical parameters derived using a NH$_3$ abundance ratio $\chi_{\text{NH}_3}=1\times10^{-8}$ and using a CS abundance ratio $\chi_{\text{CS}}=4\times10^{-9}$. }}\\
    \multicolumn{7}{p{12cm}}{\footnotesize{$^{c}$:H$_{2}$ mass and density from CS and NH$_3$ have been computed assuming the observed region is spherical or ellipsoid and whose size are given in table \ref{fitlabel} and table \ref{fitlabelbis}. }}\\
    \multicolumn{7}{p{12cm}}{\footnotesize{$^{d}$:H$_{2}$ mass are derived using a X$_{\text{CO}}=2.0\times10^{20}$cm$^{-2}$ (K km/s)$^{-1}$ and assuming a spherical region.}}\\
    \multicolumn{7}{p{12cm}}{\footnotesize{$^{e}$:Virial mass is computed using CO(1-0) and $^{13}$CO(1-0) emission FWHM and assuming a spherical region. The left value represents the Virial mass for a 1/$r^2$ density distribution whereas the right value indicate the value for a Gaussian distribution.}}\\
    \end{tabular}

\end{subtables}
\end{minipage}
\end{table*}


\clearpage
\setcounter{table}{0}
\begin{table*}
\centering
\begin{minipage}{\textwidth}
\begin{subtables}
\setcounter{table}{4}
\footnotesize
\centering
\caption{Parameters derived towards region \textit{R5} using the molecular transition CS(1-0), CO(1-0), $^{13}$CO(1-0), NH$_{3}$(1,1), NH$_3$(2,2) and C$^{34}$S(1-0) when emission are found.
    The labels \textit{a},\textit{b},\textit{c} denote a distinct emission. Lower limits have been derived when we assumed $\tau=0$ }
\begin{tabular}{ccccccccc}
    \toprule
 CS &R5 & distance & $\frac{W_{\text{CS}\left(1-0\right)}}{W_{\text{C}^{34}\text{S}\left(1-0\right)}}$ & $\tau_{\text{CS}\left(1-0\right)}$  & $N_{\text{CS}}$ [10$^{12}$]$^{a}$ & $N_{\text{H}_2}$[10$^{20}$]$^{ab}$&  $M_{\text{H}_2}^{abc}$ & $n_{\text{H}_2}^{abc}$ \\
    &  & (kpc) & & & (cm$^{-2}$)& (cm$^{-2}$)& ($\solmass$)& (cm$^{-3}$) \\
    \midrule 
    & \multirow{2}{*}{R5a} & near : 4.0 & \multirow{2}{*}{-}  & \multirow{2}{*}{-} & \multirow{2}{*}{$>3$} &\multirow{2}{*}{$>7$} & $>7.4\times10^{2}$ & $>5.7\times10^{1}$ \\
    & & far : 12.1 & & & & & $>6.7\times10^{3}$ & $>1.9\times10^{1}$\\
    \\
    & \multirow{2}{*}{R5b} & near : 3.7 & \multirow{2}{*}{-} & \multirow{2}{*}{-} & \multirow{2}{*}{$>8$} &\multirow{2}{*}{$>21$} & $>1.8\times10^{3}$ & $>1.9\times10^{2}$ \\
    & & far : 12.4 & &  & & & $>2.1\times10^{4}$ & $>5.5\times10^{1}$ \\

    \bottomrule
    \end{tabular}
    \vspace{0.5cm}
  \begin{tabular}{ccccccccc}
  \toprule
NH$_3$ &R5 & distance & $\tau_{\text{NH}_3\left(1,1\right)}$& $T_{\text{kin}}$ & $N_{\text{NH}_3}$ [10$^{12}$] & $N_{\text{H}_2}\left[10^{20}\right]^{ab}$ & $M_{\text{H}_2}^{abc}$ & $n_{\text{H}_2}^{abc}$ \\
& & (kpc)& & (K)& (cm$^{-2}$)& (cm$^{-2}$)& ($\solmass$)& (cm$^{-3}$) \\
\midrule
 &\multirow{2}{*}{R5b}& near : 3.7 & \multirow{2}{*}{-} & \multirow{2}{*}{-} & \multirow{2}{*}{$>21$} & \multirow{2}{*}{$>21$} & $>2.0\times10^{3}$ & $>1.9\times10^{2}$ \\
& & far : 12.4 &  & & & & $>2.1\times10^{4}$& $>5.6\times10^{1}$ \\
\bottomrule

  \end{tabular}
  
  \vspace{0.5cm}

    \begin{tabular}{cccccccc}
    \cmidrule[1pt]{1-6}
   CO & R5 & distance  & $M_{\text{H}_2}^{d}$  & $\Mvir\left(^{13}\text{CO}\right)^{e}$& $n_{\text{H}_2}^{e}$ \\
  & & (kpc)  &($\solmass$) & ($\solmass$) & (cm$^{-3}$)\\
  \cmidrule{1-6}
    &\multirow{2}{*}{R5a}& near : 3.7  & $9.5\times10^{3}$& $3.8\times10^{4}-1.4\times10^{5}$& $6.8\times10^{2}$ \\
    & & far : 12.4 & $1.1\times10^{4}$ & $1.3\times10^{5}-4.6\times10^{5}$&$2.0\times10^{2}$ \\
    \\
     &\multirow{2}{*}{R5b}& near : 4.0  & $1.1\times10^{4}$& $1.5\times10^{4}-5.4\times10^{4}$&$6.1\times10^{2}$ \\
    & & far : 12.1  & $1.0\times10^{5}$& $4.6\times10^{4}-1.6\times10^{5}$&$1.9\times10^{2}$ \\
    \\
     &\multirow{2}{*}{R5c}& near : 4.5 & $3.4\times10^{3}$& $3.7\times10^{4}-1.3\times10^{5}$ & $1.4\times10^{2}$ \\
    & & far : 11.6 & $2.3\times10^{4}$ & $8.0\times10^{4}-2.9\times10^{5}$ & $5.0\times10^{1}$\\
    \cmidrule[1pt]{1-6}
    \multicolumn{8}{p{12cm}}{\footnotesize{$^{a}$:Parameters have been derived using the LTE assumption.}}\\
    \multicolumn{8}{p{12cm}}{\footnotesize{$^{b}$:The H$_{2}$ physical parameters derived using a NH$_3$ abundance ratio $\chi_{\text{NH}_3}=1\times10^{-8}$ and using a CS abundance ratio$\chi_{\text{CS}}=4\times10^{-9}$. }}\\
    \multicolumn{8}{p{12cm}}{\footnotesize{$^{c}$:H$_{2}$ mass and density from CS and NH$_3$ have been computed assuming the observed region is spherical or ellipsoid and whose size are given in table \ref{fitlabel} and table \ref{fitlabelbis}. }}\\
    \multicolumn{8}{p{12cm}}{\footnotesize{$^{d}$:H$_{2}$ mass are derived using a X$_{\text{CO}}=2.0\times10^{20}$cm$^{-2}$ (K km/s)$^{-1}$ and assuming a spherical region.}}\\
    \multicolumn{8}{p{12cm}}{\footnotesize{$^{e}$:Virial mass is computed using CO(1-0) and $^{13}$CO(1-0) emission FWHM and assuming a spherical region. The left value represents the Virial mass for a 1/$r^2$ density distribution whereas the right value indicate the value for a Gaussian distribution.}}\\
    
    \end{tabular}

\end{subtables}
\end{minipage}
\end{table*}

\clearpage

\setcounter{table}{0}
\begin{table*}
\centering
\begin{minipage}{\textwidth}
\begin{subtables}
\setcounter{table}{5}
\centering
\caption{Parameters derived towards region \textit{R5} using the molecular transition CS(1-0), CO(1-0), $^{13}$CO(1-0), NH$_{3}$(1,1), NH$_3$(2,2) and C$^{34}$S(1-0) when emission are found.
     Lower limits have been derived when we assumed $\tau=0$ }

    \begin{tabular}{ccccccccc}
    \toprule
  CS &R6 & distance & $\frac{W_{\text{CS}\left(1-0\right)}}{W_{\text{C}^{34}\text{S}\left(1-0\right)}}$ & $\tau_{\text{CS}\left(1-0\right)}$  & $N_{\text{CS}}$ [10$^{12}$]$^{a}$ & $N_{\text{H}_2}$[10$^{20}$]$^{ab}$&  $M_{\text{H}_2}^{abc}$ & $n_{\text{H}_2}^{abc}$ \\
    &  & (kpc) &  & & (cm$^{-2}$)& (cm$^{-2}$)& ($\solmass$)& (cm$^{-3}$) \\
    \midrule
    
    & \multirow{2}{*}{R6} & near : 3.7 & \multirow{2}{*}{-} & \multirow{2}{*}{-} & \multirow{2}{*}{$>30$} &\multirow{2}{*}{$>75$} & $>3.6\times10^{4}$ & $>2.2\times10^{2}$\\
    & & far : 12.5 & & & & & $>4.2\times10^{5}$ & $>6.5\times10^{1}$ \\

    \bottomrule
    \end{tabular}
\vspace{0.5cm}
\centering

    \begin{tabular}{cccccccc}
    \cmidrule[1pt]{1-6}
 CO & R6 & distance  & $M_{\text{H}_2}^{d}$  & $\Mvir\left(^{13}\text{CO}\right)^{e}$&$n_{\text{H}_2}^{e}$ \\
  & & (kpc)   &($\solmass$)  & ($\solmass$)& (cm$^{3}$)\\
  \cmidrule{1-6}
    &\multirow{2}{*}{R6b}& near : 3.7  & $7.6\times10^{4}$ & $2.2\times10^{4}-7.9\times10^{4}$& $4.3\times10^{2}$\\
    & & far : 12.5  & $8.7\times10^{5}$& $7.6\times10^{4}-2.7\times10^{5}$& $1.7\times10^{2}$ \\
    \cmidrule[1pt]{1-6}
    \multicolumn{7}{p{12cm}}{\footnotesize{$^{a}$:Parameters have been derived using the LTE assumption.}}\\
    \multicolumn{7}{p{12cm}}{\footnotesize{$^{b}$:The H$_{2}$ physical parameters derived using a NH$_3$ abundance ratio $\chi_{\text{NH}_3}=1\times10^{-8}$ and using a CS abundance ratio $\chi_{\text{CS}}=4\times10^{-9}$. }}\\
    \multicolumn{7}{p{12cm}}{\footnotesize{$^{c}$:H$_{2}$ mass and density from CS and NH$_3$ have been computed assuming the observed region is spherical or ellipsoid and whose size are given in table \ref{fitlabel} and table \ref{fitlabelbis}. }}\\
    \multicolumn{7}{p{12cm}}{\footnotesize{$^{d}$:H$_{2}$ mass are derived using a X$_{\text{CO}}=2.0\times10^{20}$cm$^{-2}$ (K km/s)$^{-1}$ and assuming a spherical region.}}\\
    \multicolumn{7}{p{12cm}}{\footnotesize{$^{e}$:Virial mass is computed using CO(1-0) and $^{13}$CO(1-0) emission FWHM and assuming a spherical region. The left value represents the Virial mass for a 1/$r^2$ density distribution whereas the right value indicate the value for a Gaussian distribution.}}\\
    
    \end{tabular}

\end{subtables}
\end{minipage}
\label{lastpage}
\end{table*}

\end{document}